# High-pressure growth effects on the superconducting properties of Sm-based oxypnictide superconductors

Mohammad Azam[1], Manasa Manasa[1], Tatiana Zajarniuk[2], Ryszard Diduszko[3,4], Taras Palasyuk[1], Tomasz Cetner[1], Andrzej Morawski[1], Cezariusz Jastrzębski[5], Andrzej Szewczyk[2], Michał Wierzbicki[5], Shiv J. Singh[1,††]

[1]*Institute of High Pressure Physics (IHPP), Polish Academy of Sciences, Sokołowska 29/37, 01-142 Warsaw, Poland*

[2]*Institute of Physics, Polish Academy of Sciences, Aleja Lotników 32/46, 02-668 Warsaw, Poland*

[3]*Łukasiewicz Research Network Institute of Microelectronics and Photonics, Aleja Lotników 32/46, 02-668 Warsaw, Poland*

[4]*National Center for Nuclear Research, ul. Andrzeja Sołtana 7, 05-400 Otwock, Świerk, Poland*

[5]*Faculty of Physics, Warsaw University of Technology, Koszykowa 75, 00-662 Warsaw, Poland*

[††]Corresponding author: Shiv J. Singh

Email: sjs@unipress.waw.pl

# Abstract

High-pressure synthesis can be an effective method for improving the sample quality of materials as well as their superconducting properties. In this paper, the synthesis process of F-doped SmFeAsO has been optimized by preparing a series of bulk SmFeAsO$_{0.8}$F$_{0.2}$ (Sm1111) using the high gas pressure and high-temperature synthesis (HP-HTS) method, considering various growth parameters like growth pressures (0-1 GPa) and heating time (0.5-2 h). Structural, microstructural, Raman spectroscopic, transport, and magnetic measurements are employed to comprehensively analyze these bulks and derive the conclusive findings. The parent SmFeAsO$_{0.8}$F$_{0.2}$ prepared by the conventional synthesis process at ambient pressure (CSP) has a transition temperature ($T_c$) of around 53-54 K, and the critical current density ($J_c$) of 10$^3$ A/cm$^2$ at 5 K with a small amount of the impurity phases (SmOF and SmAs), consistent with previous reports. Interestingly, all bulks synthesized by HP-HTS have almost the same $T_c$ and $J_c$ as the parent sample. The optimal growth conditions are obtained as 900ºC, 1 hour, and 0.5 GPa with the sealed Ta-tube, which slightly improved the sample quality and the superconducting properties compared to other bulks grown by HP-HTS. Our study confirms that the existence of the impurity phases in the 1111 family is very robust and cannot be reduced by HP-HTS, leading to only a small variation in the observed superconducting properties of Sm1111 whether prepared by CSP or HP-HTS. This is the first comprehensive investigation of the high-pressure development of Sm1111, which shows distinct behaviour from other families of iron-based superconductors.

**Keywords:** Iron-based superconductors, Hot-isostatic pressing, Transport and magnetic properties, Critical current density, Critical transition temperature

# Introduction

The identification of F-doped LaFeAsO in 2008 led to the discovery of iron-based superconductors (FBS), which have a transition temperature ($T_c$) of 26 K [1]. Following this pioneering work, many superconducting compounds have been reported, and the onset $T_c$ is reached up to 58 K [2], accompanied by high critical fields ($H_{c2}$) of 100 T and high critical current densities ($J_c$) of $10^7$-$10^8$ A/cm$^2$ at 5 K [3], [4]. These properties make FBS a strong contender for high magnetic field applications. More than 100 compounds have been reported under high $T_c$ FBS, and on the basis of the crystal structures of the parent compounds, they can be categorized into six families: 11 (FeSe), 111 (*AE*FeAs; *AE* = Li, Na), 1111 (*RE*FeAsO; *RE* = Rare Earth), 122 (*A*Fe$_2$As$_2$; *A* = Ba, K), 1144 (*AeA*Fe$_4$As$_4$; *Ae* = Ca, *A* = K), and those with thick perovskites blocking layers like 42622 (Sr$_4$V$_2$O$_6$Fe$_2$As$_2$, Sr$_4$Sc$_2$O$_6$Fe$_2$P$_2$) [4], [5], [6]. These families exhibit diverse structural and superconducting properties, making them a significant area of research in high-temperature superconductivity. In all these families, 1111 is a particularly promising and interesting family that provides the highest transition temperature of 58 K in F-doped SmFeAsO (Sm1111). This family has a tetragonal ZrCuSiAs-type structure with alternating *RE*O and FeAs layers stacked along the c-axis [7]. The *RE*O layers serve as charge reservoirs, while the FeAs layers function as superconducting layers. Although the parent compound *RE*FeAsO does not exhibit superconductivity and undergoes magnetic and structural transitions around 155 K[8], [9], chemical doping plays an important role in inducing superconductivity within this family by suppressing of the magnetic and structural transitions of the parent compound. Various dopants, such as F-doping at O-sites and Co/Mn/Ni doping at Fe sites, and Sb/P at As-sites, have been reported[10][11], [12]. Among these dopants, the most effective doping level is fluorine doping at O-sites, which significantly enhances the superconducting transition to the highest value of $T_c$ = 58 K [13][14]. The optimal doping of fluorine has been reported as 20% for various 1111 members, such as La1111, Ce1111, Sm1111 [15] [16]. One basic challenge for this interesting 1111 family is to prepare the high-quality sample without the impurity phases[17]. Almost all studies have used the conventional synthesis method at ambient pressure (CSP) to grow the 1111 family, but always there exist the impurity phases such as SmOF/Sm$_2$O$_3$ and SmAs[5], [18], [19]. These impurity phases generally reduce the superconducting properties, especially the critical current density. At high temperatures and ambient pressure, the stoichiometric ratio of the compound can change due to the evaporation of the lighter elements, such as fluorine and arsenic. Therefore, the unspecified stoichiometric ratio and the high reactivity of rare earth elements generally produce the impurity phases, which reduce the grain connectivity, the sample density, and suppress the superconducting properties. The last 16 years after the discovery of 1111 family suggest that CSP method is not enough to improve the sample quality and is demanding to apply some new techniques, such as the high-pressure growth process. High-quality and large-size single crystals of F-doped Sm1111 were not possible to grow and are still challenging by the conventional growth method. The first time, Karpinski et al. [20][21], had applied a high-pressure technique to grow 1111 single crystals, where the obtained crystals were very tiny and reproducible, having a size of ~300 μm and exhibiting the superconducting properties by fluorine and cobalt doping. Some Co-doped Sm1111 crystals were used to perform Angle-resolved Photoemission Spectroscopy (APRES) measurements especially for a particular composition $SmFe_{0.92}Co_{0.08}AsO$ [22], [23]. Anger et al. recently reported the growth of Co-doped Sm1111 single crystal at the ambient pressure and performed the various transport properties [24]. Nonetheless, the absence of three-dimensional, large-sized, and well-formed F-doped Sm1111 single crystals with varying doping levels continues to pose a considerable obstacle, hindering comprehensive investigations into their magnetic and superconducting properties. Electron doping in this 1111 family by hydrogen doping has also been reported by high-pressure techniques, revealing a double dome shape in the variation of transition with H-doping contents ($x$), with a first superconducting dome (0.05<x<0.20), and a second dome (0.2<x<0.5), which is completely different behaviour than that of fluorine doping [25], [26], [27]. These findings have provided significant motivation for further exploration of the 1111 family of FBS through the high-pressure growth method [28], [29], and more research works are needed in this direction. Furthermore, very few studies have been reported using high-pressure growth for the electron doping into 1111 bulks by fluorine [30], [31] [32] without any optimized growth processes since the discovery of FBS. Recent studies based on the high gas pressure and high-temperature synthesis (HP-HTS) method have been reported for Fe(Se,Te) (11) and $CaKFe_4As_4$ (1144) families which improve the sample quality as well as also enhance the superconducting properties [33], [34], [35], [36]. These studies motivated us to apply this HP-HTS technique for this interesting 1111 family, especially for F-doped Sm1111 which provides the highest $T_c$ of 58 K for high-temperature FBS. High-pressure growth effects of F-doped Sm1111 could be helpful for understanding and exploring its superconductivity properties as well as the sample quality.

In this study, we optimize the growth condition of F-doped Sm1111 ($SmFeAsO_{1-x}F_x$) by preparing the optimal fluorine doping content ($x$ = 0.2) bulks under various growth parameters by using the high-pressure and high-temperature synthesis (HP-HTS) method up to the growth pressure of 1 GPa. These studies are compared with the parent sample (SmFeAsO$_{0.8}$F$_{0.2}$) prepared using the conventional synthesis process at ambient pressure (CSP) to understand the effect of high-pressure growth and its superconducting properties. All samples were well characterized by the structural, microstructural, Raman spectroscopy, transport, and magnetic measurements to reach the final conclusions. The preparation of Sm1111 bulks with high superconducting properties is critically dependent on each step of the growth processes. Interestingly, several Sm1111 bulks prepared by CSP and HP-HTS have almost the same transition temperature (~53-54 K) and same order of critical current density ($J_c$) under the different growth pressure. This is completely different behaviour in comparison with other iron-based superconductors [34], [35], [36]. Careful analysis of various experiments suggests that the optimal growth conditions for F-doped Sm1111 are 900°C, 0.5 GPa, and 1 h with the sealed Ta-tube. The bulk prepared under these optimal conditions has depicted a slightly improved sample density, transition temperature, and critical current density compared to other samples. These studies suggest that the observed impurity phases in F-doped Sm1111 prepared either by HP-HTS or CSP are very robust, causing almost no changes in its superconducting properties.

## Experimental details

Bulk SmFeAsO$_{0.8}$F$_{0.2}$ was prepared by the single step solid-state reaction method. Samarium (Sm) powders (99.9%), iron (Fe) powders (99.99%), arsenic (As) chunks (99.99%), iron oxide (Fe$_2$O$_3$) powder (99%), and iron fluoride (FeF$_2$) powder (99%) were used as the initial precursors. Since the element "As" is highly reactive as a pure element, hence we prefer to prepare "SmAs" by mixing Sm and As powder, and after pelletizing, it was sealed in an evacuated quartz tube, which was heated at 550 ºC for 15 hours. This prepared SmAs powder was used as a precursor. In the next steps, all these precursors are mixed according to the stoichiometric chemical formula of SmFeAsO$_{0.8}$F$_{0.2}$ having a weight of 6-7 grams in one batch. After thoroughly mixing of this powder, 5-6 disk pellets having a diameter of 11 mm were prepared by hydraulic pressing at 200 bars. These pellets were placed inside a Ta-tube and then sealed inside an evacuated quartz tube, which was heated at 900 ºC for 45 hours [37]. All the synthesis process was performed inside the high-purity glove box with very low level (< 1 ppm) of oxygen and moisture. One can note that we have used the nominal composition of $SmFeAsO_{0.8}F_{0.2}$ in our study.

The high-pressure synthesis of F-doped Sm1111 was conducted using the high gas pressure and high temperature synthesis (HP-HTS) technique [34]. An inert gas pressure of up to 1.8 GPa can be generated in a cylinder chamber utilizing a three-zone furnace capable of achieving temperatures up to 1700 °C [38]. A programmable temperature controller can precisely manage the temperature stability to within 0.5 K, and this pressure approach offers many cm3 of free volume for the sample's growth. High isostatic gas pressures of up to 1.0 GPa for a range of response periods have been achieved in all our studies using the argon gas atmosphere. Along with a three-stage gas compressor, our experimental apparatus contains an internal single- or three-zone furnace in our high-pressure technical chamber. After the first step of reaction using CSP, as discussed above, these pellets are used in two ways: ground up and pelletized, and direct pellet of the parent sample P, which is described in Figure 1. To avoid any deterioration during the growing phase, we conducted all chemical manipulations in a glove box filled with argon. Several samples from various batches were prepared using this method, and each one showed excellent consistency in terms of superconducting characteristics.

To analyze the high-pressure growth effect, a series of $SmFeAsO_{0.8}F_{0.2}$ was prepared by CSP, as a first step. In the second step, the growth process was started by grinding the parent sample P into a fine powder to apply high-pressure growth, as marked as G-batch (Figure 1). We prepared several pellets, each weighing approximately one gram, from this powder. Each pellet was sealed individually inside a tantalum (Ta) tube using an ARC Melter. In total, five Ta-tubes were prepared for this experiment. In the first batch denoted as G-batch, one Ta tube was heated at 900°C for 1 hour under ambient pressure (0 GPa), resulting the sample labelled as G0. The other four Ta-tubes were subjected to varying pressures of 0.3 GPa, 0.5 GPa, 0.7 GPa, and 1 GPa, yielding samples G1, G2, G3, and G4, respectively. All samples were synthesized at a constant temperature of 900 °C for 1 hour. This procedure allowed us to investigate the effects of increasing pressure on the growth and properties of F doped Sm1111. For the second batch, denoted as the T-batch, the samples were synthesized at a fixed temperature of 900 °C and a pressure of 0.5 GPa, but at different heating times, as shown in Figure 1. We subjected sample T1 to 0.5 hours, T2 to 1 hour, and T3 to 2 hours which allowed us to study the impact of different growth durations on the resulting material. The third batch, denoted as the D-batch, involved the direct use of the parent sample without grinding. The sample was sealed in a Ta-tube and subjected to high-pressure conditions at a constant temperature of 900 °C for 1 hour. Sample D0 was synthesized at ambient pressure (0 GPa), while samples D1 and D2 were synthesized at 0.5 GPa and 1 GPa, respectively. The pressure and temperature conditions for all three batches (G, T, and D) are summarized in Figure 1, which shows the graphical representation of the growth parameters used in each set of experiments. Following the HP-HTS growth process, the resulting pellets were compact in nature, yet not in the form of disks. The samples have been meticulously shaped into rectangular bars to facilitate microstructural analysis, Raman spectroscopy, resistivity, and magnetic measurements. To facilitate a rapid assessment of the heating temperature, a sample was prepared at 600 °C for 1 hour under 0.3 GPa pressure, with its characterizations illustrated in the supplementary figures S1, S2, and S3.

Room temperature X-ray diffraction (XRD) using a Rigaku Smart Lab 3 kW diffractometer was used to characterize the structural properties of the bulk materials. With a step rate of 0.01°/min, the measuring profile of this instrument ranged from 20° to 70° using filtered Cu-Kα radiation (wavelength: 1.5418 Å, power: 30 mA, 40 kV). The gathered diffraction data were analyzed utilizing the ICDD PDF4+ 2024 standard diffraction patterns database alongside Rigaku's PDXL software to identify impurity phases and lattice properties. A Zeiss Ultra Plus field-emission scanning electron microscope (FESEM) equipped with a Bruker Quantax 400 EDS microanalysis system was utilized to perform an in-depth microstructural analysis and elemental mapping. The magnetic measurements of these samples are conducted using a vibrating sample magnetometer (VSM) linked to a Quantum Design Physical Property Measurement System (PPMS), operating within a temperature range of 2-60 K and under a magnetic field of 9 T. The magnetic susceptibility ($\chi$) was assessed in both zero-field-cooled (ZFC) and field-cooled (FC) modes under an applied magnetic field of 20 Oe. The transport properties, specifically the resistivity, were assessed utilizing a four-probe technique with a closed-cycle refrigerator (CCR) across a temperature range of 6-300 K in the absence of a magnetic field. The data were collected via a slow warming procedure to study the temperature dependence of resistivity under different current ($I$ = 0.1, 1, 10 and 20 mA).

We have performed Raman scattering measurements by using the LabRam ARAMIS spectrometer from Horiba Jobin Yvon. Samples were stimulated using visible light at a wavelength of 632.8 nm from a He-Ne ion laser. Light was concentrated using a 100x objective with a numerical aperture of 0.95, achieving a focal spot of less than 3 μm on the examined sample. The backscattered light collected from the same objective was subsequently dispersed by a diffraction grating of 2400 l/mm and recorded by a charge-coupled device (CCD), which provides the range of the Stokes portion of Raman shifts from 65 to 300 cm$^{-1}$. Laser power was attenuated to a few milliwatts (mW) to avoid sample overheating. Raman spectra were acquired at room temperature with an acquisition time of 120 s or longer, averaged by 3 accumulations. The sample surface was measured as provided post-synthesis, without any further treatment applied. The collected spectra regarding both the intensity and position of the detected peaks are considerable scattered due to the used polycrystalline samples. These observations may be attributed to varying grain orientations and lattice strains resulting from F substitution. Raman spectra were collected for each content of F substitution, averaging data from at least ten points on the sample to ensure sufficient statistics regarding peak positions. We have deconvoluted the measured spectra by considering a Lorentzian shape.

## Results and discussion

### 1. Structural analysis

The XRD pattern of the parent bulk (P) prepared using the CSP process is shown in Figure 2(a). It shows tetragonal ZrCuSiAs-type structure and has a small amount of SmOF and SmAs phases, as listed in Table 1. The observed lattice parameters are $a$ = 3.929(9) Å and $c$ = 8.502(6) Å and the lattice volume $V$ = 131.0 Å$^3$ as like the previous reports [39], [40], [41], [42]. XRD patterns for all G batch samples are shown in Figure 2(a) with the parent P. Interestingly, all these sample G1, G2, G3 and G4 have the tetragonal phase and the observed impurity phases of SmOF and SmAs as like G0 and parent P samples. Interestingly, these impurity phases are varied very small with the applied growth pressure, as mentioned in Table 1. Careful analysis suggests that the sample G2 has the lowest amount of impurity phases compared to other samples (G0, G1, G3-G4) and almost like the parent P. The enlarged view of the main peak (102) of the superconducting phase is shown in Figure 2(b), which depicts a slight shift of peak position towards higher angles for the sample prepared at the applied pressure, i.e. G1, G2, G3 and G4. It suggests a slight decrease in the lattice parameters for these samples compared to the sample G0 prepared at 0 GPa. The calculated lattice parameters '$a$', '$c$' and the unit cell volume '$V$' are depicted in Figure 2(c)-(e). Figure 2(c) displays a consistent value for the lattice parameter '$a$' for all samples, whereas the lattice parameter '$c$', see in Figure 2(d), is reduced with the applied growth pressure, which agrees well with the shift of the peak position as shown in Figure 2(b). The unit cell volume ($V$), Figure 2(e), is slightly higher for G0 compared to P, but seems slightly reduced for the samples G1, G2, G3 and G4 compared to G0. The increased $V$ of the sample G0 at zero pressure could be due to a small evaporation of fluorine, but the applied growth pressure reduces this evaporation, and the cell volume is reduced for G1-G4. It suggests the enhancement of fluorine contents inside the superconducting layer due to the applied growth pressure. Hence, these studies recommend that the Sm1111 sample prepared by HP-HTS after the grinding and pelletized process of the parent P sample has the best growth conditions: 0.5 GPa and 1 hour at 900 °C.

In the next step, we have used the pellet of the parent sample P directly to the HP-HTS process without the grinding process to check the effect on the high-pressure growth of F-doped Sm1111. Three samples were prepared at 0, 0.5 and 1 GPa, and marked as D0, D1 and D2, as depicted in Figure 1. The high-pressure synthesis of Sm1111 by using grinding and pelletized sample, i.e. a series of sample G0-G4, as discussed above, has suggested the best pressure of 0.5 GPa for the growth for Sm1111. Following it, we have decided to consider three pressures for the direct use of parent P pellet, i.e., 0, 0.5 and 1 GPa, and their powder XRD patterns are shown in Figure 3(a). All these samples have tetragonal phase and consist of similar impurity phases (SmOF, SmAs), as discussed above, and mentioned in Table 1. The enlarge view of the main 102 peak, Figure 3(b), has almost the same peak position as the parent P and D0 samples. However, the samples D0 and D1 have a slight larger amount of impurity phases (especially SmAs) compared to the parent P sample, and it is increased for the high growth pressure, i.e. 1 GPa, the sample D2. The calculated lattice parameters $a$, $c$, and the lattice volume $V$ are depicted in Figure 3(c)-(e). The lattice parameters '$a$' and '$c$' are almost constant for D0, D1, and D2, as shown in Figure 3(c) and (d). Furthermore, the lattice volume 'V' for these samples displays almost no changes (Figure 3(e)) and suggests that the direct use of pellets P during high pressure growth slightly enhanced the impurity phases, as confirmed from Figure 3(a) and Table 1. However, its effects on the lattice parameters and the lattice volume $V$ are very small to detect for D0, D1, and D2. It proposes that the actual content of F is slightly reduced from the superconducting layer. Compared to the sample G2 and D1 prepared at 0.5 GPa, it seems that the growth conditions for the sample G2 are more suitable for the 1111 phase formation.

To optimize the heating time, we have prepared F-doped Sm1111 bulks by grinding and palletizing process of parent P at the applied growth pressure of 0.5 GPa and different heating times of 0.5 h, 1 h, and 2 h. The samples are marked as T1, T2 and T3, as mentioned in Figure 1, and the rest of the processes were the same as those of the G-series samples. The powder XRD patterns for these samples are shown in Figure 4(a). All samples have tetragonal phase and similar impurity phases of SmOF and SmAs as observed for the parent P sample. The amount of SmOF impurity phase is ~10% for T1, ~6% for T2 and ~11% for T3 sample. So, the impurity phases for T2 are very much similar to the parent P. The enlarged view of the 102 peak is almost in the same position for all T-batch samples with the parent P, as depicted in Figure 4(b). Due to the slightly increased amount of SmOF phase for the samples T1 and T3, it is sure that the actual amount of F contents is reduced from the superconducting phase of Sm1111. However, the change of actual F contents is very small and inhomogeneous, so we could not observe a clear shift of the main 102 peak position, as depicted in Figure 4(b). The calculated lattice parameters $a$, $c$, and volume $V$ are shown in Figure 4(c)-(e). There is a very small enhancement in the lattice parameters '$a$' and '$c$' for the samples T1 and T3 compared to the sample T2, as observed from Figure 4(c) and (d). The unit cell volume $V$ reaches to minima for the T2 sample, which suggests slightly an increment in the actual fluorine contents for F-doped Sm1111 bulks compared to T1 and T3. These analyses suggest that a small amount of fluorine is reduced for long time heated samples at the growth pressure of 0.5 GPa. On this basis, we can conclude that long heating time for the preparation of Sm1111 is not suitable and increased the amount of the impurity through the HP-HTS process, which is like the reports based on the CSP method [43]. All these studies based on the structural analysis suggest that the bulk G2 prepared at 0.5 GPa and 1 hour are the best F doped Sm1111 grown by HP-HTS compared to other samples.

## 2. *Raman spectroscopy measurements and analysis*

Raman scattering of all three G, D and T- batch samples are investigated in the study and revealed three major spectral features: signals related to Sm(A1g), As(A1g), and Fe(B1g) out-of-plane lattice modes, as shown in Figure 5(a). That was in line with earlier observations reported in the literature [37], [44][45][46]. In general, Raman spectra collected for each sample from at least 2 probing points showed very slight scatter in values of peak position (less than 1.5 cm$^{-1}$). In contrast, when comparing samples from different batches, experimentally determined phonon energies showed considerably higher degree of value scatter falling within 2 cm$^{-1}$, 4 cm$^{-1}$ and 1 cm$^{-1}$ spectral range for Sm, As and Fe sublattice vibrations for G, D and T-batch samples, as depicted in Figures 5(b), (c), and (d) respectively. Considering the accuracy of determined values (shown as vertical error bars), we could not plausibly trace any correlation between the energy of measured phonons and the change of "$c$" lattice parameter (observed in XRD measurements presented here, especially, for samples from "G" batch).

Nonetheless, it should be emphasizing that this was observed for polycrystalline samples, which might be significantly affected by internal intergrain stresses. The effect of compression along the "*c*" direction on phonon energy can be further investigated either experimentally by carrying out Raman scattering measurements on single crystalline samples or by theoretically assessing (DFT calculations) phonon energies for unit cells with lattice parameters experimentally determined in this study for polycrystalline samples. One may speculate that a small reduction of the unit cell along the "*c*" dimension determined from the XRD could be a consequence of a small increase in the actual fluorine doping contents in Sm1111 under high-pressure synthesis. Smaller atoms of Fluorine occupy positions of oxygen atoms that should lead to decreasing of the layered unit cell, particularly, in the "*c*" crystallographic direction. Due to this, ones should have a definitive effect on the energy of lattice vibrations involving atoms of Samarium that occupy the neighbouring atomic layer along the "*c*" direction in the unit cell. As already mentioned, we noticed a slight but definitive softening of A1g Raman mode of Sm sublattice in the samples of "G" series. The reason for this could be followed by comparing the energies of Sm–O and Sm–F bonds. Further DFT calculations should definitively shed light on the possible mechanism behind the observed changes in XRD and Raman measurements.

3. *Elemental mapping and analysis*

To collect the elemental mapping and its analysis, these samples were polished by the sandpaper with different grades without using any lubricant. We have performed the elemental analysis mapping on these samples using the energy dispersive X-ray (EDAX) method, which allows us to understand the distribution of the constituent elements inside the bulk sample. Figure S4, as the supplementary figure, displays the elemental mapping of the samples G0, G1, G2, G3 and G4, prepared under various pressures of 0, 0.3, 0.5, 0.7, and 1 GPa, respectively, relative to the parent sample P. Figure S4(i) confirms the almost homogeneous distribution of the constituent elements Sm, Fe, As, O and F. At very few places, the rich areas of Sm, As, O and F are observed, which suggests the presence of a small amount of SmAs and SmOF/$Sm_2O_3$ phase, as like the XRD analysis. The mapping for the sample G0 is shown in Figure S4(ii), which looks very much like the parent sample P and confirms the presence of the same impurity phases. The ambient growth pressure has a homogeneous distribution of Sm1111 and the impurity phases inside the whole sample area, where the amount of impurity phases is observed constant for the sample P and D0. The sample prepared at the applied pressure of 0.3 GPa, i.e. G1, has a large area of inhomogeneous distribution of the constituent elements, suggesting the accumulation of the impurity phase of SmAs and SmOF in a large area where the rest parts of the sample are homogeneous with the Sm1111 phase, as shown in Figure S4(iii). This suggests that the low growth pressure led to the accumulation of impurity phases in small areas. Further increase of the growth pressure, i.e. 0.5 MPa, sample G2 results in the improved homogeneity of the constituent elements, but the accumulated area of impurity phases is decreased compared to the sample G1 and increased the homogeneity of the superconducting Sm1111 area. The Sm1111 samples prepared at 0.7 GPa, i.e. G3, have again increased the accumulated area of the impurity phases compared to G0, and G2, but similar to the sample G1 as depicted in Figure S4(iv). Hence, this growth pressure is not very supportive to the formation of the superconducting 1111 phase, which is like the case of 0.3 GPa i.e. sample G1. For higher growth pressure, i.e. 1 GPa, the mapping of the sample G4 is shown in Figure S4(vi). There are many large and small areas observed for the impurity phases due to the accumulation effect, and the amount of impurity phases seems slightly increased compared to the samples G2, G0 and P. Low growth pressure, i.e., 0.3 GPa and high pressure, i.e., 0.7-1 GPa are not very supportive for the phase formation of Sm1111 and lead to accumulation of the impurity phases in the large areas, i.e, inhomogeneity of the constituent elements. Overall, we can say that the applied pressure of 0.5 GPa used for G2 from the G-batch could be an optimal growth pressure for the homogeneity of the Sm1111 phase.

The elemental mapping for the samples D0, D1, and D2 is illustrated in Figure S5 as the supplementary data, which is performed using the direct pellet of the parent P sample at the applied pressures of 0, 0.5, and 1 GPa. The sample D0 has an almost homogeneous distribution of the constituent elements Sm, As, Fe, F, and O, but in a few places, SmAs and SmOF are observed, as for the parent sample P. The mapping for the sample D1, prepared at 0.5 GPa, is shown in Figure S5(ii). The impurity phases (SmOF and SmAs) are observed in many larger areas in the whole sample. The observed impurity phases seem to increase compared to the sample D0. Hence, it shows that the direct pellet processed sample at 0.5 GPa has enhanced the inhomogeneity of the constituent elements i.e., amount of the impurity phases, but the impurities are distributed in the whole sample area. Further increase of the growth pressure of 1 GPa, i.e., the sample D2, inhomogeneous distribution of elements is observed in the large area of the sample, supporting the accumulation of the impurity phases in large areas. However, the amount of impurities are slightly increased compared to the samples D1 and D0, as mentioned in Table 1. The results for the D batch suggest that high pressure growth of Sm1111 accumulates the impurity phases in the large areas of the sample with a slight increase in the amount of impurity phases (Table 1). These analyses suggest that the use of direct pellets of the parent P is not suitable for high pressure growth of F-doped Sm1111, and even the sample D1 prepared at 0.5 GPa has a large inhomogeneous distribution of the constituent elements compared to that of the sample G2.

The elemental mapping of the samples prepared at different heating times, i.e. T1, T2 and T3, are shown in Figure S6 as the supplementary data file. The sample T1 prepared at 0.5 h has an almost homogeneous distribution of Sm, Fe, As, O and F, Figure S6(i), but at some places, the impurity phases (SmAs or SmOF) are detected. The observed areas with respect to SmAs seem large and observed at many places. The distribution of the constituent elements in the sample T2, Figure S6(ii), has many small-sized inhomogeneous areas throughout the whole sample and suggests the existence of SmAs or SmOF/$Sm_2O_3$ phase as like the xrd analysis. Further increasing the heating time has accumulated the impurity phases in many large areas, as depicted in Figure S6(iii) for the sample T3. This effect is very much similar to the preparation of Sm1111 bulks at high growth pressure, i.e., for the sample G3 and G4. The long heating time seems supporting the enhancement of the impurity phases in large areas. The analysis of the sample T1 and T3 suggests the slightly large amount of impurity phases compared to the sample T2, as listed in Table 1. Hence, the long heating time under the growth pressure of 0.5 GPa also promotes the formation of large areas of the impurity phases, where its amount does not change too much but accumulates in the large area. Based on these mapping analyses of G-series, D-series, and T-series, the optimal conditions for the growth of F doped Sm1111 by HP-HTS should be 0.5 GPa, 1 h at 900°C. This analysis is in good agreement with the XRD measurements as discussed above.

## *4. Microstructural analysis*

For the microstructural analysis, the polished samples (Parent P and G, D and T-batches) by several grade of emery paper are used for the scanning electron microscope, and many images in backscattered electron (BSC) mode are collected with various magnifications. One can note that BSE images reveal the chemical contrast as well as the microstructure of bulk samples. Figure 6(a-r) depicts BSE images of various samples of G-batch with various growth pressures. The images have three contrasts: bright, light grey and black corresponding to $Sm_2O_3$/SmOF, SmFeAsO$_{0.8}$F$_{0.2}$, and pores, respectively. The black contrast can be SmAs/FeAs at few places. The parent P has many small grains, and there do exist nanopores between many grains, which reduce the grain connections and suggest the low sample density, as shown in Figure 6(a)-(c).

Also, the impurity phases $Sm_2O_3$/SmOF and SmAs are distributed homogeneously in the whole sample area (Figure 6(a)), as well agreed with the above-discussed elemental mapping. After grinding and pelletizing the parent P, the sample G0 prepared at 0 GPa has a large area of $Sm_2O_3$ phase at many places, and SmAs are also observed, as depicted in Figure 6(d-e), but the rest of the sample seems homogeneous with the superconducting phase 1111. Like the parent P, many small grains and nanopores in between superconducting grains do exist. It suggests that many grains are not well connected, suggesting low sample density. For the sample G1, the areas of the impurity phase are increased by reducing the number of small areas as observed for G0, and grain connections seems like the samples G0 and P. Basically, the impurity phase is accumulated in a few large areas, and the rest of the sample looks homogeneous with the Sm1111 phase, as also suggested by the elemental mapping. Further increase of the growth pressure to 0.5 GPa, i.e., the G2 sample, the areas of impurity phase $Sm_2O_3$ are reduced, and observed at some places, but in a few places, the SmAs phase is also observed. However, a large part of the sample looks homogeneous and dense compared to the samples G0, G1, and P, which suggests the better grain connectivity than these bulks. The area of impurity phase again starts to increase for the higher growth pressure of 0.7 GPa (sample G3) and 1 GPa (sample G4). Interestingly, the sample G3 has many small grains which have poor intergrain connections. For 1 GPa, i.e. the sample G4, the impurity phases (SmAs and $Sm_2O_3$) are observed at many small areas in the whole sample. So, it is expected the reduced sample density compared to the sample G2 but probably should be the same as that of the sample G0 and P. It seems that the growth pressure of 0.5 GPa seems a suitable pressure; lower or higher than this pressure enhances the area of impurity phases in the whole sample. But, on average, the amount of impurity phases seems almost the same for all the samples G0-G4 as that of the parent P sample. By considering the theoretical density of 7.1 g/cm$^3$ for Sm1111, the calculated sample density is listed in Table 1, which suggests that the sample G2 has an improved density of 58% compared to other samples, which agrees well with the microstructural analysis. The analysis of G series samples suggests that the sample G2 grown under the pressure of 0.5 GPa has the best growth conditions, which is supported by the XRD and elemental mapping outcomes.

For the samples grown using the direct pellet of the parent P sample, i.e., D-series, BSE images are shown in Figure S7(a-l), as the supplementary figure. The parent sample P is also included in this figure to make an easy comparison to D-series samples. Figure S7(d)-(f) for the sample D0 confirms the presence of the impurity phases and pores. The micropores do exist, and in a few places, well-connected grain boundaries are observed. In the sample D0, the impurity phases of $Sm_2O_3$ are observed in a smaller area compared to the sample G0, and the SmAs phase exists at many places. However, the grain connectivity seems almost the same as the parent P and G0 samples. BSE images of the sample D1 show the impurity $Sm_2O_3$ and SmAs phases at many places as the same as the parent P, as depicted in figures S7(g)-(i). There do exist some well-connected grain boundaries and micro/nanopores in between grains (Figure S7(h)-(i)). At few places, some small cracks are also observed. Further increase of the growth pressure, i.e., the sample D2, the impurity phase, especially $Sm_2O_3$ phase, is accumulated in large areas, but the rest of the sample looks homogeneous with the superconducting Sm1111 phase. Many pores and long cracks are also observed in the sample D2, as shown in Figure S7(l). The analysis of BSE images of these samples suggests that the same impurity phases do exist for all these samples and the high growth pressure reduces the grain connectivity due to the presence of cracks and the impurity phases. Also, it seems that the impurity phase SmAs are observed slightly higher than the G batch. Table 1 lists the calculated sample density of the samples D0, D1 and D2 using the theoretical density of 7.1 g/cm$^3$ for Sm111. On can see that on average, there is almost no improvement in the sample density due to the HP-HTS process compared to that parent P, which agrees well with the microstructural analysis. Hence, the direct use of pellets of the parent P does not improve the sample density or reduce the presence of impurity phases. The samples G2 and D1 have the same growth conditions, Figure 1, except one sample was ground and pelletized, and the other one is a direct pellet of the parent P, but BSE analysis of these two samples suggests that the sample G2 is better than that of the sample D1. Hence, the direct use of pelletized parent P is not very supportive to improve the sample quality even at the same growth pressure.

For third batch samples i.e., T-series, BSE images are depicted in Figure S8 (supplementary data file) with respect to the parent P. The parent P has a homogeneous distribution of the impurity phases of SmAs and $Sm_2O_3$ in the whole sample as many small areas. The sample T1 has many large accumulated area of the impurity phase $Sm_2O_3$ and nano/micropores exist at many places, as shown in Figure S8(d)-(f). The impurity phase SmAs is also observed at many places as small areas, which is same as the parent P. On this basis, the grain connectivity of the sample T1 can be expected to be the same as the parent P (Figure S8(a)-(c)). The sample T2 heated for 1 hour has reduced the area of the impurity phase $Sm_2O_3$ and looks more compact than the sample T1 and P, as shown in Figure S8(g)-(i). Micro/nano pores in between grains are slightly reduced at this heating time, and the sample has a more homogeneous distribution of the Sm1111 phase compared to other samples. For further increase of heating time to two hours, i.e., the sample T3, a large accumulated area of the impurity phases ($Sm_2O_3$ and SmAs) is observed (Figure S8(j)); however, the rest part of the sample has the superconducting Sm1111. There do exist many nano/micropores (Figure S8(k)-(l)). Some well-connected grain boundaries are also observed for this sample T3. The presence of impurity phases and pores reduced the grain connections and the effect of high-pressure growth. Table 1 presents the sample density calculated for Sm1111 using the theoretical density of 7.1 g/cm$^3$. Comparison all these samples from the T-series suggests that the sample T2 has fewer pores, a lower impurity phase, a higher sample density, and a homogeneous distribution of the superconducting Sm1111 phase compared to the samples T1 and T3. Hence, both low and long heating times are not good for the growth of Sm1111 phase. This study confirms that the heating time of 1 hour at the growth condition of 900°C and 0.5 GPa is most suitable to improve the sample quality with homogeneous distribution of Sm1111 phase, there is no need to apply the high growth pressure i.e., more than 0.5 GPa. This is well in agreement with the above-discussed analysis of the structural and elemental mapping.

### 5. *Transport properties*

The resistivity behaviours of G-batch Sm1111 bulks prepared by HP-HTS using the growth pressure of 0-1 GPa are illustrated in Figure 7(a) in the temperature from 7 K to 300 K. The parent P sample has a room temperature resistivity of ~5.5 mΩ-cm at room temperature, which decreases linearly with temperature and shows the superconducting transition at 53.5 K. It is like the previous reported F-doped Sm1111 [11], [47]. The reground and pelletized parent sample P was heated at 900 °C, for 1 h at zero pressure, i.e., the sample G0 exhibits a linear resistivity behaviour and is shown in Figure 7(a). Interestingly, the resistivity value is increased almost two times in comparison with the parent sample P, which could be due to the observed many large areas of $Sm_2O_3$ phase or inhomogeneity of the constituent elements and slightly reduced sample density, as discussed above. The sample prepared under the applied pressure of 0.3 GPa, i.e., the sample G1 has reduced resistivity in the whole temperature range, which is slightly lower than the parent P and much lower than that of the sample G0. This behaviour well agrees with the microstructure of the sample G1 and suggests the improvement of grain connectivity and homogeneity of the constituent elements compared to G0. For further increase of the growth pressure up to 0.5 GPa, i.e., the sample G2, further reduction of the resistivity is observed that suggests the improvement of the grain connections and sample density, as listed in Table 1. The sample G3 prepared at 0.7 GPa has slightly higher resistivity than that of G2, which could be due to the accumulation of the impurity phase, especially $Sm_2O_3$ in the larger area and the reduction of the sample density. The sample G4 prepared at the growth pressure of 1 GPa has much lower resistivity in comparison to all other samples (P, G0, G1-G3); even this sample has a slightly large amount of impurity compared to other samples (G0-G3, P), and the sample density is the same as that of the parent P. It can be understood by considering the effects of this high growth pressure effect on the clean parts of the sample. The normal state resistivity behaviour of all these samples has similar behaviour as that of the parent sample P and the previous reports [37], [39]. To be clearer, the low temperature resistivity behaviours for these samples are shown in Figure 7(b) in the temperature range of 30 K to 60 K and compared with the parent P sample. The sample G0 has the onset transition temperature ($T_c^{onset}$) of 52 K, which is slightly improved with 0.3 and 0.5 GPa. The sample G0 has not reached zero resistivity down to 7 K, which confirms the presence of a large amount of non-superconducting phase and weak link behaviours. The enhancement of transition temperature could be due to the slight improvement of the actual fluorine contents in the sample, as indicated by XRD data analysis. For further increase of pressure to 0.7 and 1 GPa, the $T_c^{onset}$ is almost the same value of ~53 K. The broadening of the superconducting transition is reduced with the applied growth pressure, suggesting the improvement of the grain connections. The grain connections are improved by the high-pressure growth at the clean parts of the sample, but due to the presence of the same amount of impurity phases, the combined effects are observed as the average effects of the whole sample.

For the D-batch samples, the temperature dependence of resistivity for different samples D0, D1 and D2 is depicted in Figure 8(a) in the temperature range 7 to 300 K. Both samples D0 and D1 have almost the same resistivity behaviour, like the parent sample P, but their resistivity values are much lower than the parent sample P, which could be possible due to a slight improvement in the sample density or grain connections. The resistivity decreases linearly down to the superconducting transition. Due to its very high resistivity value, the resistivity behaviour of D2 is shown in the inset of Figure 8(a). The large resistivity value could be due to the presence of large cracks in this sample D2 compared to the samples D0 and D2, as discussed in Figure S7. To understand the superconducting behaviour, the low temperature resistivity of these samples is shown in Figure 8(b). The onset $T_c$ value is almost the same (~ 53 K) for these samples D0, D1 and D2. However, the broadening of the transition temperature for the sample D1 is reduced compared with other samples prepared at 0.5 or 1 GPa. Large transition broadening ($\Delta T$) suggests weak grain connections, which could be due to the presence of a large area of $Sm_2O_3$ as observed from the microstructural analysis. The growth pressure of 0.5 GPa has reduced the area of $Sm_2O_3$ and SmAs as observed in D0 and D2 (Figure S7), which slightly improved the sample density and grain connections. Therefore, these analyses based on D-series suggest a suitable growth pressure of 0.5 GPa, supporting the output from D-series. These findings well agreed with the microstructural and structural analysis.

The temperature dependence of resistivity for the samples T1-T3 prepared under different heating times at a growth pressure of 0.5 GPa is shown in Figure 9(a). The sample heated at 0.5 h, i.e., T1 has shown the transition below 200 K, which is like the without F-doped SmFeAsO [8], [48] and has a large resistivity value at room temperature. From the microstructural analysis, this sample has many large areas of $Sm_2O_3$ (Figure S8) and suggests that the large part of the sample is not superconducting. Hence, the resistivity behaviour of the sample T1 is well agreed with the microstructural analysis. The sample T2 heated at 1 h at 0.5 GPa has resistivity reduced almost 10 times in comparison with the sample T1 and has lower resistivity than the parent P sample. For further increase of heating time, i.e., for the sample T3, the resistivity is reduced compared to the samples T1, T2 and P. However, the behaviour of T2 and T3 samples is the same as that of the parent P. The low temperature behaviour of these samples is shown in Figure 9(b). The sample T1 shows a drop-in resistivity around 53 K, but the zero resistivity is not observed. For the sample T2, the observed onset $T_c$ value is ~53-54 K, and zero resistivity is observed at ~43 K. Furthermore, the sample T3 has the onset $T_c$ that is slightly lower than that of T2 and has a very broad transition width, as shown in the inset of Figure 9(b). This analysis suggests that the heating time of 1 hour at 0.5 GPa is the best condition for the growth of the superconducting 1111 phase.

## 6. *Magnetic properties:*

To verify the Meissner effect for these samples, we performed dc magnetic susceptibility measurements in both zero-field-cooled (ZFC) and field-cooled (FC) mode in the temperature ranges 5-60 K and the applied magnetic field of 20 Oe for the selected samples. We have normalized the magnetization curve by the magnetization at 5 K to make a comparative study of these bulk samples. Figure 10(a) depicts the normalized magnetization curve for the parent sample P, G2 and G4. The observed relatively small value for FC magnetization suggests the strong vortex pinning in the bulk samples and the permanent trapped inside it. The samples G2 and G4 have almost the same onset transition temperature as that of the parent sample P. A two-step transition is observed more clearly for the sample G4 compared to G2, which is same as that of the parent P sample and suggests a weak link behaviour as reported for FBS [4]. The temperature dependence of the normalized magnetization for the samples D1 and D2 is shown in Figure 10(b) together with the parent P sample. The sample D1 has the same onset $T_c$ of 53 K, like the resistivity transition. The parent P sample has a two-step transition, while one-step transition is observed for D1, indicating the improved grain connections for F doped Sm1111 at 0.5 GPa. For further increase of high-pressure growth, i.e., 1 GPa, the sample D2 has the almost same onset $T_c$ but also depicts paramagnetic impurity phases in the normal state, which could be due to the presence of an unknown phase in the xrd data (Figure 3(a)) and well agreed with the above discussed microstructure and structural analysis. The samples with different heating times at 0.5 GPa are shown in Figure 10(c) for the samples T2 and T3. Both samples (T2 and T3) have the onset $T_c$ of ~53 K, whereas the sample T3 depicted two-step transition and a separation of the ZFC and FC curve in the normal state. This could be possible due to the presence of FeAs phase at a few places in this sample T3 (Figure S6(iii)), as observed in the microstructural analysis. These magnetic susceptibility measurements also confirm that the heating time of 1 h at 0.5 GPa is more suitable compared to other heating time. It supports the analysis of other measurements.

For the practical application, the critical current density ($J_c$) plays an important role for a superconductor. The magnetic hysteresis (*M-H*) loops at 5 K were measured for the parent sample P and some selected samples prepared by HP-HTS from all three batches, Figure 1, under the magnetic field up to 9 T. The hysteresis loops for the parent P, G2 and G4 are shown in the inset to Figure 11(a), and their behaviour is like the previous reports [37]. The width of the hysteresis loop can be calculated by considering the magnetic moment in the ascending and descending the magnetic field. Generally, the Bean model is used to calculate the critical current density by applying the formula: $J_c = 20\Delta m/Va(1-a/3b)$ [49], where *a* and *b* are the length of the sample (*a<b*) and *V* is the volume of the sample. The rectangular-shaped samples are used for the magnetic measurements. By considering the sample dimensions, the calculated the current density is depicted in Figure 11(a). The parent compound has the $J_c$ value of the order of $10^3$ A/cm$^2$ which is slightly reduced with the magnetic field up to 9 T which is like the reported F doped Sm1111 [2, 11]. The $J_c$ is improved for the sample G2 in the whole magnetic field range. Further increase of the growth pressure reduces the $J_c$ value as observed for the sample G4. It could be due to the presence of impurity phase or inhomogeneity in the sample, as observed from the structural and microstructural analysis. The pinning force has a direct impact on the increment of the critical current density. To examine the pinning force characteristics of these samples, we have determined the relationship between the applied magnetic field and the density of vortex pinning force, $F_p$, using the equation $F_p = \mu_0 H \times J_c$. This calculated pinning force is depicted in Figure S9(a) in the supplementary data file for the sample P, G2 and G4. The pinning force analysis confirms the enhancement of the $J_c$ is due to the increment of pinning force compared to the parent sample P.

For the next batch of samples, the temperature dependence of $J_c$ is shown in Figure 11(b) for the selected samples D1 and D2. The inset of figure 11(b) depicts the *M-H* loop for these samples. The results show the enhancement of $J_c$ for the sample D1, which is prepared at 0.5 GPa. Further increase of the pressure reduce the $J_c$ value as like the sample G4. The enhancement of $J_c$ suggests the increment of pining force as shown in figure S9(b) (supplementary data file) compared to the parent P sample. To see the heating time effect on $J_c$, the calculated $J_c$ value and the corresponding pinning force are shown in Figure 11(c) and S9(c). *M-H* loop of these samples are depicted in the inset of Figure 11(c). The obtained $J_c$ of T2 exhibits the same the behaviour as the parent P sample but its value is slightly increased in the whole magnetic field. Long-time heated sample, i.e., the sample T3 has reduced the critical current density compared to the sample T2 due to the presence of the cracks and has same magnetic field dependence of the sample P and T2. The obtained pinning force is shown in Figure S9(c), which suggests that the enhancement of $J_c$ supports the increase of pinning force as like other samples. These analyses advise that the heating time of 1 hour is the best condition to improve the critical current density and pinning force of F doped Sm1111 at 0.5 GPa through HP-HTS. The reduced $J_c$ value and flux pinning centres could be due to the reduced sample density or the presence of impurity phases.

## 7. *Discussion:*

To summarize the findings from high pressure synthesis by the HP-HTS method, we have plotted the onset transition temperature $T_c^{onset}$, transition width ($\Delta T$), the room temperature resistivity ($\rho_{300K}$), Residual Resistivity Ratio ($RRR = \rho_{300\,K}/\rho_{60\,K}$) and the critical current density ($J_c$) for these three batches of samples in Figure 12, 13 and 14. The onset transition is 52 K for the sample G0, which is slightly increased with the growth pressure and reached to a maximum value of 53 K for the sample G2, i.e. the synthesis pressure of 0.5 GPa (Figure 12(a)). Further increase of the growth pressure, $T_c^{onset}$ starts to decrease, as observed for the samples G3 and G4. Interestingly, the broadening of transition width is also reduced linearly with the growth pressure up to 0.7 GPa and after that it is almost constant, as shown in Figure 12(b). This suggests that the high-pressure growth effect improves the grain connection. The room-temperature resistivity is also reduced by the high-pressure growth process, as shown in Figure 12(c). The calculated *RRR* for these samples is revealed in Figure 12(d), which is increased linearly with the growth pressure. The pressure ranges from 0.3 to 0.7 GPa i.e., G1-G3, has an almost constant value of RRR, which is increased rapidly for the growth pressure of 1 GPa. It suggests that the overall the growth pressure improves the homogeneity and grain connectivity of F-doped Sm1111. Figure 12(e) depicts the critical current density ($J_c$) at 0.2 and 8 T at 5 K for G2 and G3 prepared at 0.5 GPa and 1 GPa. The obtained critical current density is higher than the parent P prepared by CSP. This analysis confirms that the reground and pelletized the parent P sample prepared by the HP-HTS process at 900 °C, 1 h has the best growth pressure of 0.5 GPa compared to others growth pressures. These analyses suggest that the high-pressure growth improves the grain connectivity behaviours and homogeneity of the samples, especially superconducting regions, by accumulating the impurity phase in a large area, as discussed above with the microstructural analysis.

To understand the high-pressure effect for the direct pellet of the parent P without grinding, the extracted various parameters are shown in Figure 13. The onset transition temperature $T_c^{onset}$ is 53.2 K for the D0 sample (Figure 13(a)), which is almost constant for the samples D1 and D2 under the pressure growth of 0.5 and 1 GPa, respectively. Interestingly, the transition is very broad for the sample D0, as shown in Figure 13(b). The transition width ($\Delta T$) is decreased very rapidly for the sample D1 grown under the pressure of 0.5 GPa. Further increase of the growth pressure, i.e., the sample D2, again increased the broadening of the transition very rapidly, which reflects the weak grain connections due to the presence of impurity phases, the reduced sample density or inhomogeneity of the samples (Table 1). The variation of room temperature resistivity ($\rho_{300K}$) is illustrated in Figure 13(c). The samples D0 and D1 have almost the same resistivity value, which could be due to the almost same sample density and impurity phases (Table 1). However, it increases rapidly for the high growth pressure, i.e., the D2 sample, which could be due to the presence of large cracks as observed from the microstructural analysis. The RRR is also shown in Figure 13(d), which is almost same for D0 and D1, but it reduced for high growth pressure of 1 GPa, i.e., the sample D2. The reduced RRR advises the inhomogeneity of the superconducting phase due to the accumulation of the impurity phases in many large areas, as discussed above. The calculated $J_c$ of the sample D1 at 0.2 T and 8 T is higher than the sample D2. The study of these bulks under the direct batch has also confirmed that 0.5 GPa is a more suitable growth pressure for the F-doped SmFeAsO phase, which is well agreed with the grinded batch samples (G-series) as discussed above with Figure 12. More than 0.5 GPa pressure produced a large crack inside the bulks and degraded the sample's quality very rapidly.

The heating time dependence of the $T_c^{onset}$ for T-batch samples is shown in Figure 14(a). The samples heated for 0.5 h and 1 h have the almost same onset $T_c$ value of 53 K, like the parent P sample. However, for a long-heated samples i.e., T2 sample, the transition temperature is reduced, which could be due to the evaporation of fluorine for this sample, as discussed above. The variation of transition width $\Delta T$ with heating time is shown in Figure 14(b). The sample T1 does not reach to zero resistivity, so not included in this figure. The sample T2 has almost the same transition width as like that of the parent P. The $\Delta T$ is enhanced for the T3 bulk, reflecting the weak grain connections compared to the sample T2. The variation of room temperature resistivity, $\rho_{300\ K}$, with heating time is shown in Figure 14(c). It decreases very rapidly from 0.5 h to 1-hour heating. Further increments of heating time slightly reduced the room temperature resistivity as observed for the sample T3. The *RRR* value is almost same for the sample T1 and T3, whereas the sample T2 heated for 1 hour has a slightly lower value, as depicted in Figure 14(d). It could be possible due to the reduced impurity phases and the rest of the sample is homogeneous with the superconducting phase compared to T1 and T3. The critical current density $J_c$ at 5 K for T2 and T3 is shown in Figure 14(e). The sample T2 has shown a large $J_c$ value at low and high magnetic field compared to the sample T3. The outcomes from Figure 14 suggest that a heating time of 1 hour is most favourable for the conditions of 900 ºC and 0.5 GPa for the synthesis of the 1111 phase.

Finally, to make a clearer understanding, the growth pressure dependence of $T_c^{onset}$, the critical current density $J_c$ are depicted in Figure 15(a)-(b) for all these bulks grown by the HP-HTS process. From Figure 15(a), the samples from the G-batch (G0-G4) clearly depict the highest $T_c^{onset}$ value for the sample G2, i.e., for the growth pressure of 0.5 GPa. The samples (D0-D2) from the direct batch show a slightly lower $T_c^{onset}$ for D1, prepared at 0.5 GPa. Furthermore, T-batch samples are prepared under the different heating times, where T2 heated at 1 h has a higher $T_c^{onset}$ value than the sample T3. One can note that the T1 sample, as depicted in Figure 15(a), has the onset $T_c$, but the zero resistivity is not reached. In Figure 15(b), the samples G2/T2 from the ground and pelletized batch have a higher $J_c$ value than the sample G4, whereas the sample D1 from the direct batch D has a higher $J_c$ value than D2 but less than G2 and T2. The samples T2 and T3 are also included in Figure 15(b), in which the sample T2 has shown a higher $J_c$ value than that of the T3 sample. To better understand of the effect of growth pressure on critical current density ($J_c$), we have plotted the amount of superconducting (SC) Sm1111 phase calculated from XRD measurements and the sample density with respect to $J_c$ value in Figures 15(c) and (d), where the parent P sample is included with samples G2, G4, D1, D2, and T3. The SC phase of G2 exhibits a significant improvement compared to other bulks, suggesting a reduction of the impurity phases and the enhancement of $J_c$ is observed. It implies that the high-pressure growth effects on the formation of Sm1111 require the right conditions. Figure 15(d) illustrates the enhancement of $J_c$ for the high density of the sample, i.e., the G2 has a higher density and $J_c$ than that of other samples. These studies suggest that the higher sample density and pure superconducting Sm1111 phase are needed to improve the $J_c$ value. The existence of the impurity phases (SmOF and SmAs) inside the grains or in between grains can reduce the $J_c$ value. Therefore, it is very important to enhance the superconducting phase and sample density during the high-pressure growth process. Hence, the optimal growth pressure of 0.5 GPa is the best pressure under the growth conditions of 900 °C, 1 h. However, these growth conditions are not sufficient to completely reduce the impurity phases and further studies are needed in this direction. Furthermore, the prepared F-doped Sm1111 at 600°C, 1 h and 0.3 GPa has a large amount of impurity phases, as confirmed from Figures S2 and S3 in the supplementary data files. Due to this, the superconducting transition of this sample is very broad and showed two steps-like behaviours, as shown in Figure S1. It suggests that this temperature of 600°C is not suitable for high pressure growth compared to the used heating temperature of 900°C. Our finding clearly recommends that a growth pressure of 0.5 GPa is enough to prepare the improved sample quality Sm1111 bulks with the slightly enhanced the critical current density. High-pressure growth of F-doped Sm1111 is not able to reduce the impurity phases even up to the applied growth pressure up to 1 GPa, and interestingly, the amount of impurity phases is almost the same as that of the sample grown at ambient pressure. Because of this, we have not observed a significant improvement in either the sample quality or the superconducting properties of F doped Sm1111 by the HP-HTS process. These results depict a different effect of high-pressure growth compared to the previous reports based on the FeSe$_{0.5}$Te$_{0.5}$ (11) and CaKFe$_4$As$_4$ (1144) [35], [36], where the superconducting properties, sample quality, and density are improved. The results and analysis presented in this manuscript necessitate further research under the high growth process to further improve the sample quality and enhance the superconducting properties of the 1111 family.

## Conclusions

We investigated the superconducting properties of bulk SmFeAsO$_{0.8}$F$_{0.2}$ (Sm1111) synthesized by CSP and HP-HTS methods. A series of bulks were prepared under the growth pressure up to 1 GPa and various conditions, and characterized them by various measurements. Interestingly, high pressure growth could not reduce the impurity phases, which were observed almost the same amount as the parent sample. The structural and microstructural analyses confirm that using grounded, palettized, and enclosed in Ta-tube Sm1111 bulks is an efficient way of increasing sample density and enhancing grain connectivity compared to other growth conditions. The resistivity and magnetization measurements suggest that a growth pressure of 0.5 GPa is enough for the growth of 1111 phase, however, a pressure lower or higher than 0.5 GPa or a long heating time promote the impurity phases, which lower the sample density and superconducting properties. Analysis of various Sm1111 prepared by HP-HTS suggests the best growth conditions: 900 °C, 1 h, and 0.5 GPa for the fluorine doping at O-sites, which has a nearly three times higher (~3.2 ×10$^3$ A/cm$^2$) $J_c$ value than the parent compound. It could be related to the improved grain connections, increased sample density, and additional pinning centres. Our studies confirm that the optimal conditions for F doped Sm1111 are 0.5 GPa and 1 h by HP-HTS, as consistent with the previous reported paper based on other FBS families. However, the high-pressure growth was not able to reduce the amount of impurity phase for F doped Sm1111, which is completely different behaviour than other FBS families. We believe, our study is useful for the development of the magnetic applications of 1111, especially for the superconducting wires and tapes. Further investigations or optimization of the growth parameters will be needed in this research direction.

## Declaration of competing interest

The authors declare that they have no known competing financial interests or personal relationships that could have appeared to influence the work reported in this paper.

**Acknowledgments:**

The work was funded by SONATA-BIS 11 project (Registration number: 2021/42/E/ST5/00262) sponsored by National Science Centre (NCN), Poland. SJS acknowledges financial support from National Science Centre (NCN), Poland through research Project number: 2021/42/E/ST5/00262.

# References

[1] Y. Kamihara, T. Watanabe, M. Hirano, H. Hosono, "Iron-Based Layered Superconductor La[O$_{1-x}$F$_x$]FeAs (x = 0.05–0.12) with $T_c$ = 26 K," *J. Am. Chem. Soc.,* **130**, 3296, 2008.

[2] S. J. Singh et al., " Transition Temperature and Upper Critical Field in SmFeAsO$_{1-x}$F$_x$ Synthesized at Low Heating Temperatures," *IEEE Trans. Appl. Supercond.,* **23**, 7300605, 2013.

[3] S. J. Singh and M. Sturza, "Bulk and Single Crystal Growth Progress of Iron-Based Superconductors (FBS): 1111 and 1144," *Crystals,* **12**, 20, 2022.

[4] H. Hosono, A. Yamamoto, H. Hiramatsu and Y. Ma, "Recent advances in iron-based superconductors toward applications," *Materials Today,* **21**, 278-302, 2018.

[5] J. Paglione and R. L. Greene, 'High-temperature superconductivity in iron-based materials', *Nature Physics* **6**, 645 (2010). doi: 10.1038/nphys1759.

[6] J. I. Shimoyama, 'Potentials of iron-based superconductors for practical future materials', *Supercond. Sci. Technol.* **27**, 044002, 2014.

[7] P Quebe, L. J. Terbuchtë, W. J. Jeitschkö, and J. Jeitschkö, 'Quaternary rare earth transition metal arsenide oxides *RT*AsO (*T*=Fe, Ru, Co) with ZrCuSiAs type structure', *Journal of Alloys and Compounds* **302** , 70–74, 2000.

[8] C. Hess *et al.*, 'The intrinsic electronic phase diagram of iron-oxypnictide superconductors', *EPL*, **87**, 17005, 2009

[9] R. H. Liu *et al.*, 'Anomalous transport properties and phase diagram of the FeAs-based SmFeAsO1-xFx superconductors', *Phys Rev Lett*, vol. 101, no. 8, Aug. 2008, doi: 10.1103/PhysRevLett.101.087001.

[10] S. J. Singh, J.-I. Shimoyama, A. Yamamoto, H. Ogino, and K. Kishio, 'Effects of phosphorous doping on the superconducting properties of SmFeAs(O,F)'*Physica C*, **15**, 19-23, 2014.

[11] M. Azam *et al.*, 'Antimony Doping Effect on the Superconducting Properties of SmFeAs(O,F)', *IEEE Transactions on Applied Superconductivity*, vol. **34**, no. 3, pp. 7300405 (2024), doi: 10.1109/TASC.2023.3343328.

[12] G. Lamura, T. Shiroka, S. Bordignon, S. Sanna, M. Moroni, R. De Renzi, P. Carretta, P. K. Biswas, F. Caglieris, M. Putti, S. Wurmehl, S. J. Singh, J. Shimoyama, M. N. Gastiasoro, and B. M. Andersen, "Role of magnetic dopants in the phase diagram of Sm 1111 pnictides: The case of Mn," *Phys. Rev. B,* **94**, 214517 , 2016.

[13] S J Singh *et al*., Weak-link behaviour observed in iron-based superconductors with thick perovskite-type blocking layers' Supercond. Sci. Technol. **26**,105020, 2013.

[14] X. C. Wang *et al.*, 'Revisiting the Electron-Doped SmFeAsO: Enhanced Superconductivity up to 58.6 K by Th and F Codoping', *Chinese Physics Letters*, **34**, 077401, 2017.

[15]  J. Prakash, S. J. Singh, G. Thakur, S. Patnaik, and A. K. Ganguli, 'The effect of antimony doping on the transport and magnetic properties of Ce(O/F)FeAs', *Supercond Sci Technol*, **24**, 125008, 2011.

[16] S. J. Singh, J. Prakash, S. Patnaik, and A. K. Ganguli, 'Enhancement of the superconducting transition temperature and upper critical field of LaO$_{0.8}$F$_{0.2}$FeAs with antimony doping', *Supercond Sci Technol*, **22**, 4/045017, 2009.

[17] Z. A. Ren and Z. X. Zhao, 'Research and prospects of iron-based superconductors', *Adv. Mater.* **21**, 4584 (2009) . doi: 10.1002/adma.200901049.

[18] K. Iida, J. Hänisch and A. Yamamoto, "Grain boundary characteristics of Fe-based superconductors," *Supercond. Sci. Technol.* ,**33**, 043001, 2020.

[19] Q. Si, R. Yu and E. Abrahams, "High-temperature superconductivity in iron pnictides and chalcogenides," *Nature Reviews Materials,* **1**,16017, 2016.

[20] J. Karpinski *et al.*, 'Single crystals of LnFeAsO1-xFx (Ln = La, Pr, Nd, Sm, Gd) and $Ba_{1-x}Rb_xFe_2As_2$: Growth, structure and superconducting properties', *Physica C: Superconductivity and its Applications*,**469,** 370–380, 2009.

[21] N. D. Zhigadlo, S. Katrych, Z. Bukowski, S. Weyeneth, R. Puzniak, and J. Karpinski, 'Single crystals of superconducting SmFeAsO1-xFy grown at high pressure', *Journal of Physics Condensed Matter*, **20**, 342202, 2009.

[22] A. Charnukha *et al.*, 'Interaction-induced singular Fermi surface in a high-temperature oxypnictide superconductor', *Sci Rep*, 5,10392 ( 2015).

[23] A. Charnukha *et al.*, 'High-temperature superconductivity from fine-tuning of Fermi-surface singularities in iron oxypnictides', *Sci Rep*, **5**, 18273, (2015)

[24] F. Anger *et al.*, 'On the growth and characterization of SmOFe1−xCoxAs single crystals', *J Cryst Growth*, **627**, 127473. 2024.

[25] H. Hosono and S. Matsuishi, 'Superconductivity induced by hydrogen anion substitution in 1111-type iron arsenides', Current Opinion in Solid State and Materials Science,**17**, 49–58 (2013).

[26] K. Kobayashi *et al.*, 'Pressure effect on iron-based superconductor LaFeAsO1-x Hx: Peculiar response of 1111-type structure', *Sci Rep*, **6**, 39646, 2016.

[27] H. Takahashi *et al.*, 'Superconductivity at 52 K in hydrogen-substituted LaFeAsO1-xHx under high pressure', *Sci Rep*, 5, 07829, (2015).

[28] M. Miura *et al.*, 'Quadrupling the depairing current density in the iron-based superconductor SmFeAsO1–xHx', *Nat Mater* **23**, 1370 (2024), doi: 10.1038/s41563-024-01952-7.

[29] S. Iimura, T. Muramoto, S. Fujitsu, S. Matsuishi, and H. Hosono, 'High pressure growth and electron transport properties of superconducting $SmFeAsO_{1-x}H_x$ single crystals', *Journal of Asian Ceramic Societies*, **5**, 357–363, 2017.

[30] F. Kametani *et al.*, 'Combined microstructural and magneto-optical study of current flow in polycrystalline forms of Nd and Sm Fe-oxypnictides', *Supercond Sci Technol*, **22**, 015010 (2009).

[31] R. Zhi-An *et al.*, 'Superconductivity at 55 K in Iron-Based F-Doped Layered Quaternary Compound Sm[O 1−x F x ]FeAs ', *chin. phys. lett.* **25**, 2215 (2008.

[32] Z. A. Ren *et al.*, 'Superconductivity in the iron-based F-doped layered quaternary compound Nd[O1 - X Fx]FeAs', *EPL*, **82** 57002 (2008).

[33] M. Azam, M. Manasa, A. Morawski, T. Cetner, and S. J. Singh, 'High Gas Pressure and High-Temperature Synthesis (HP-HTS) Technique and Its Impact on Iron-Based Superconductors', *Crystals (Basel)*, vol. 13, no. 10, Oct. 2023, doi: 10.3390/cryst13101525.

[34] M. Manasa *et al.*, 'Effect of impurity phase and high-pressure synthesis on the superconducting properties of CaKFe4As4', Journal of Physics and Chemistry of Solids **190**, 111996, 2024

[35] M. Manasa *et al.*, 'Enhancement of Superconducting Properties of Polycrystalline CaKFe4As4 by High-Pressure Growth', *IEEE Transactions on Applied Superconductivity*, vol. **34**, 345821, 2024.

[36] M. Azam, M. Manasa, T. Zajarniuk, R. Diduszko, T. Cetner, A. Morawski, J. Więckowski, A. Wiśniewski, S. J. Singh, "High-Pressure Synthesis and the Enhancement of the Superconducting Properties of $FeSe_{0.5}Te_{0.5}$," *Materials* **16**, 5358 (2023)

[37] M. Azam, M. Manasa, T. Zajarniuk, T. Palasyuk, R. Diduszko, T. Cetner, A. Morawski, C. Jastrzebski, M. Wierzbicki, A. Wiśniewski, S. J. Singh, "Copper doping effects on the superconducting properties of Sm-based oxypnictides, "*J. Am. Ceram. Soc*. **107** (2024).

[38] M. Azam, M. Manasa, T. Cetner, A. Morawski, S. J. Singh, "High Gas Pressure and High-Temperature Synthesis (HP-HTS) Technique and Its Impact on Iron-Based Superconductors," *Crystals* **13**, 1525 (2023).

[39] J. B. Anooja, P. M. Aswathy, P. M. Sarun, and U. Syamaprasad, 'A novel low temperature synthesis route for $SmFeAsO_{1-x}F_x$ bulk superconductor with improved transport properties', Journal of Alloys and Compounds **514**, 1–5, 2012.

[40] M. Fujioka, S. J. Denholme, M. Tanaka, H. Takeya, T. Yamaguchi, and Y. Takano, 'The effect of exceptionally high fluorine doping on the anisotropy of single crystalline $SmFeAsO_{1-x}F_x$', *Appl Phys Lett*, vol. 1**05**, 102602, 2014.

[41] B. Ni *et al.*, 'Critical current characteristics and history dependence in superconducting SmFeAsOF bulk' *Journal of Physics: Conference Series* **234**, 012028, 2010.[42] Y. Qi, Z. Gao, L. Wang, D. Wang, X. Zhang, and Y. Ma, 'Superconductivity in co-doped SmFeAsO', *Supercond Sci Technol*, **21**, 115016, (2008).

[43] M. Azam *et al.*, 'Optimization of Synthesis Parameters and Superconducting Properties of GdFeAsOF', *IEEE Transactions on Applied Superconductivity*, vol. **34**, 3341858, (2024).

[44] C. Marini, C. Mirri, G. Profeta, S. Lupi, D. Di Castro, R. Sopracase, P. Postorino, P. Calvani, A. Perucchi, S. Massidda, G. M. Tropeano, M. Putti, A. Martinelli, A. Palenzona and P. Dore, "The optical phonon spectrum of SmFeAsO," *Europhysics Letters,* **84**, 67013, (2008).

[45] L. Zhang, P. F. Guan, D. L. Feng, X. H. Chen, S. S. Xie, M. W. Chen, "Spin-Dependent Electron−Phonon Interaction in SmFeAsO by Low-Temperature Raman Spectroscopy," *J. Am. Chem. Soc,* **132,** 15223–15227, (2010).

[46] V. G. Hadjiev, M. N. Iliev, K. Sasmal, Y.-Y. Sun, C. W. Chu, "Raman spectroscopy of RFeAsO (R_Sm, La)," *Phy. Rev. B,* **77**, 220505R, (2008).

[47] S. J. Singh, J. I. Shimoyama, A. Yamamoto, H. Ogino, and K. Kishio, 'Significant enhancement of the intergrain coupling in lightly F-doped SmFeAsO superconductors', *Supercond Sci Technol*, **26**, 065006, (2013).

[48] A. Srivastava *et al.*, 'Magnetotransport and thermal properties characterization of 55 K superconductor $SmFeAsO_{0.85}F_{0.15}$', *AIP Adv*, 3, 092113. (2013).

[49] C. P. Bean et al., "Magnetization of high-field superconductors," *Rev. Mod. Phys.,* **36**, 31, 1985

**Table 1:** List of the sample code, the amount of impurity phases SmOF and SmAs from the XRD measurements, the calculated sample density.

| Sample code | SmOF (%) | SmAs (%) | Sample density (%) |
|---|---|---|---|
| P | ~7 | ~2 | 50 |
| G0 | ~8 | ~2 | 48 |
| G1 | ~10 | ~3 | 49 |
| G2/T2 | ~6 | ~2 | 58 |
| G3 | ~11 | ~3 | 51 |
| G4 | ~13 | ~3 | 50 |
| D0 | ~8 | ~9 | 51 |
| D1 | ~6 | ~10 | 51 |
| D2 | ~6 | ~14 | ~49 |
| T1 | ~10 | ~8 | 52 |
| T3 | ~11 | ~9 | 47 |

**Figure 1:** The preparation conditions of three distinct batch SmFeAsO$_{0.8}$F$_{0.2}$ samples using the HP-HTS technique under varying conditions are depicted. These batches are defined by considering the ground and direct pellet of the parent sample P.

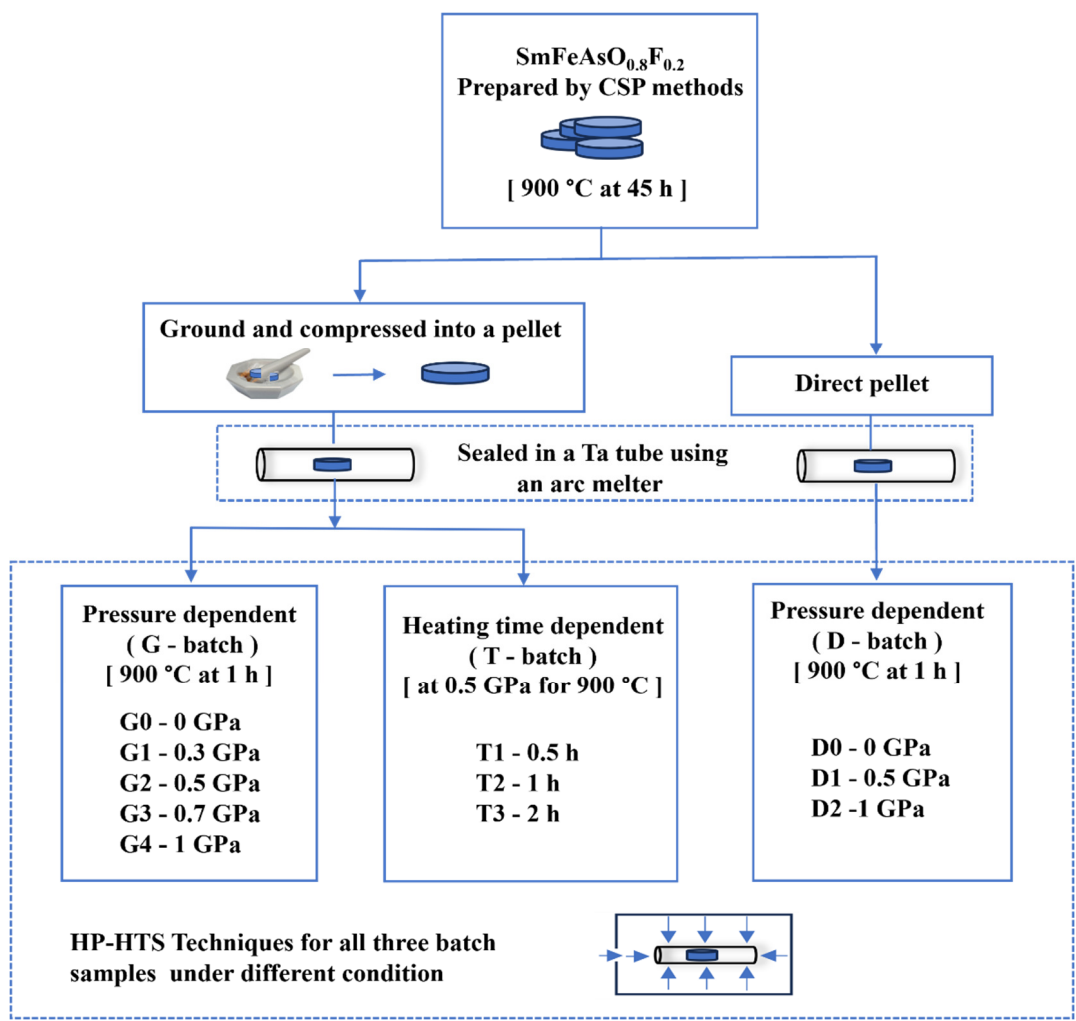

**Figure 2:** **(a)** X-ray diffraction patterns (XRD) of powder SmFeAsO$_{0.8}$F$_{0.2}$ sample (P) prepared by CSP at ambient pressure, ground, and pelletized samples G0, G1, G2, G3, and G4 prepared by HP-HTS at constant heating temperature 900 °C and time 1 h under different growth pressures of 0, 0.3, 0.5, 0.7 and 1 GPa. **(b)** An enlarged view of the main peak (102) position is depicted. The variation of **(c)** lattice parameter ($a$), **(d)** lattice parameter ($c$), and **(e)** unit cell volume ($V$) with the various synthesis pressures (0 – 1 GPa) for the sample G0, G1, G2, G3, and G4.

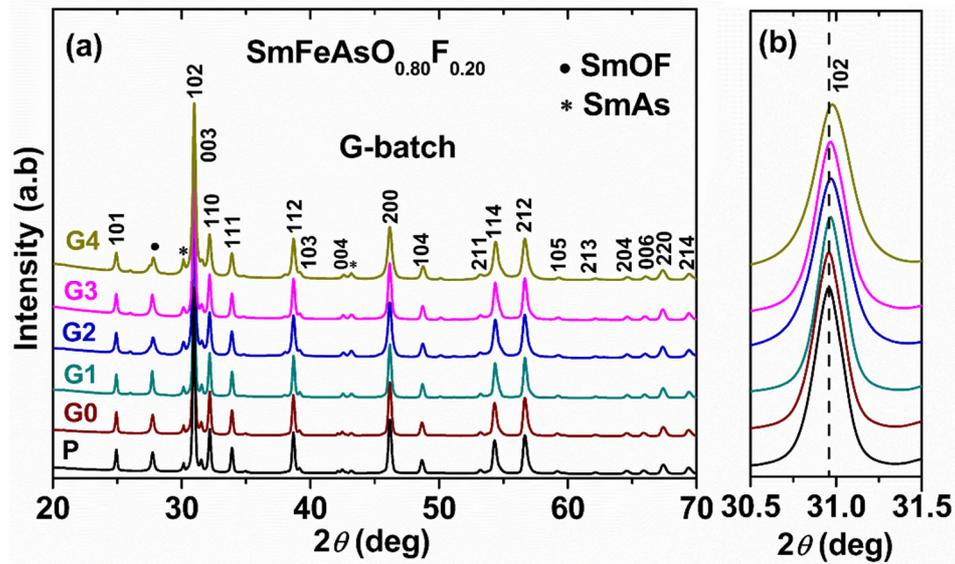

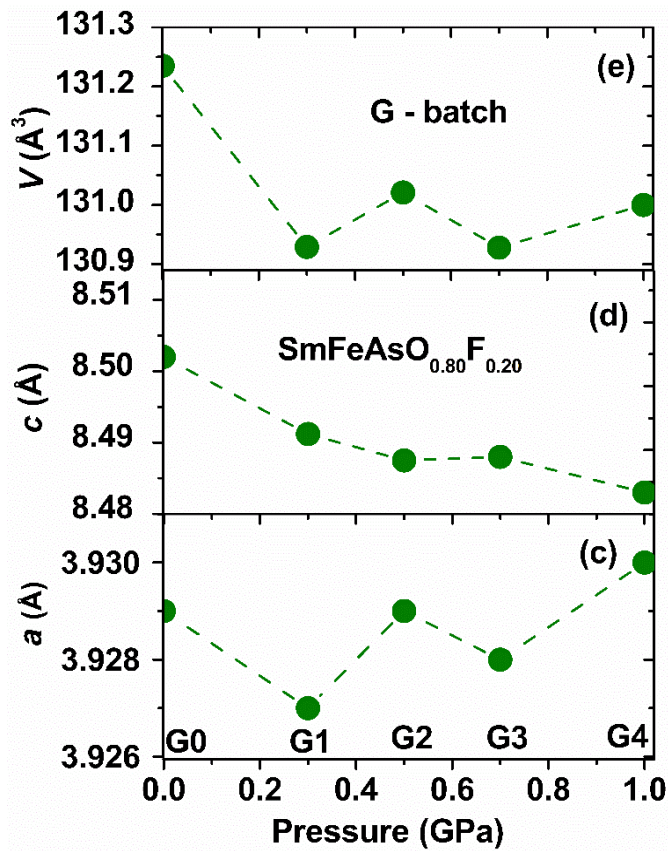

**Figure 3:** (**a**) X-ray diffraction patterns (XRD) of powder SmFeAsO$_{0.8}$F$_{0.2}$ sample of the parent P prepared by CSP at ambient pressure, and direct pellet samples D0, D1, and D2 prepared by HP-HTS at constant temperature 900 °C and time 1 h under different growth pressure of 0 GPa, 0.5 GPa, and 1 GPa. (**b**) An enlarged view of the main peak (102) position is depicted. The variation of (**c**) lattice parameter (*a*), (**d**) lattice parameter (*c*), and (**e**) unit cell volume (*V*) with the various pressures (0 – 1 GPa) for the sample D0, D1, and D2.

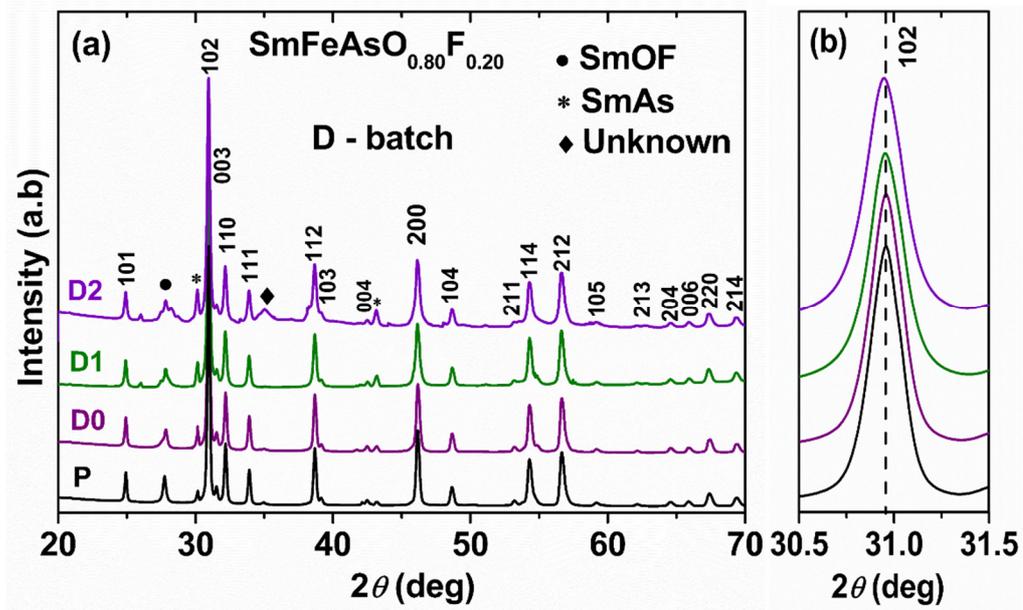

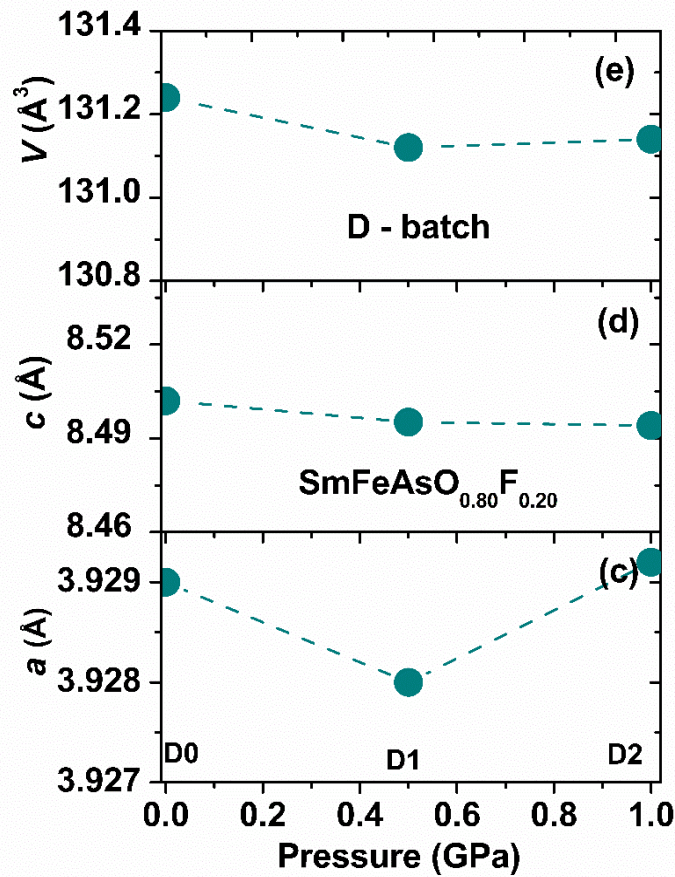

**Figure 4:** **(a)** X-ray diffraction patterns (XRD) of powder SmFeAsO$_{0.8}$F$_{0.2}$ parent P sample prepared by CSP at ambient pressure, and ground and pelletized samples T1, T2, and T3 prepared by HP-HTS at constant temperature 900 °C and the growth pressure of 0.5 GPa under different heating time of 0.5, 1 and 2 hours. **(b)** An enlarged view of the main peak (102) position is depicted. The variation of **(c)** lattice parameter (*a*), **(d)** lattice parameter (*c*), and **(e)** lattice volume (*V*) with the various growth time (0.30 – 2 h) for the sample T1, T2, and T3.

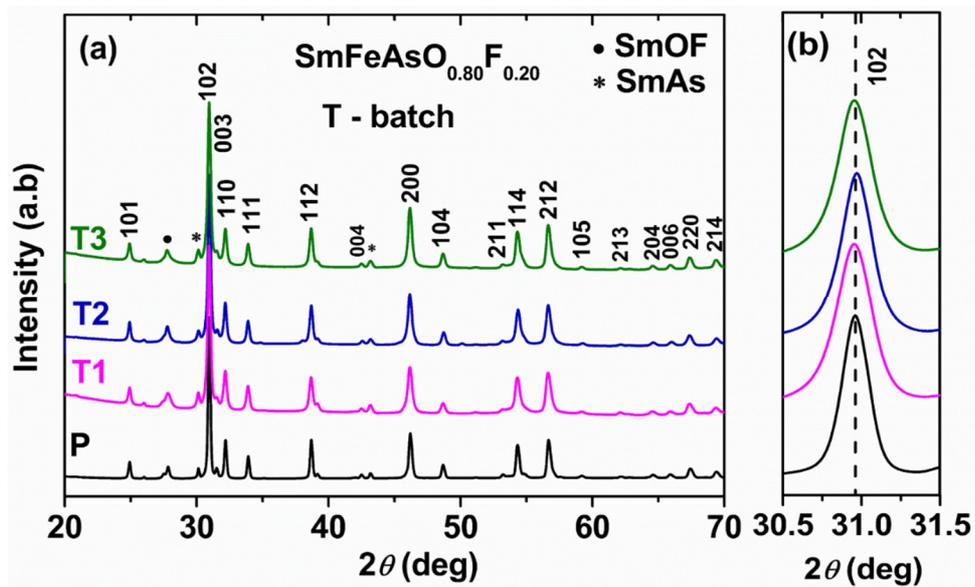

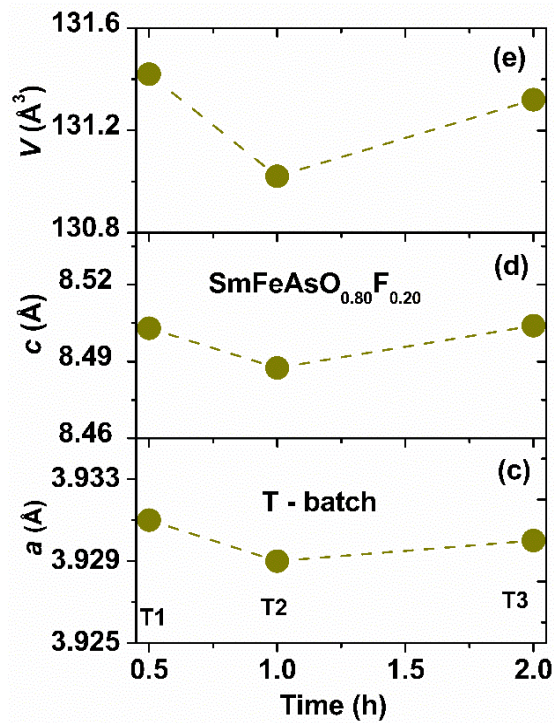

**Figure 5: (a)** Raman scattering spectrum acquired for the "G" batch is shown, and major spectral features are present in spectra obtained under different synthesis pressures. Experimental spectrum (open circles) is fitted by peaks of Lorentzian line shape (solid lines). **(b), (c), (d)** Variation of peak positions measured for samples from "G", "D" and "T" batches.

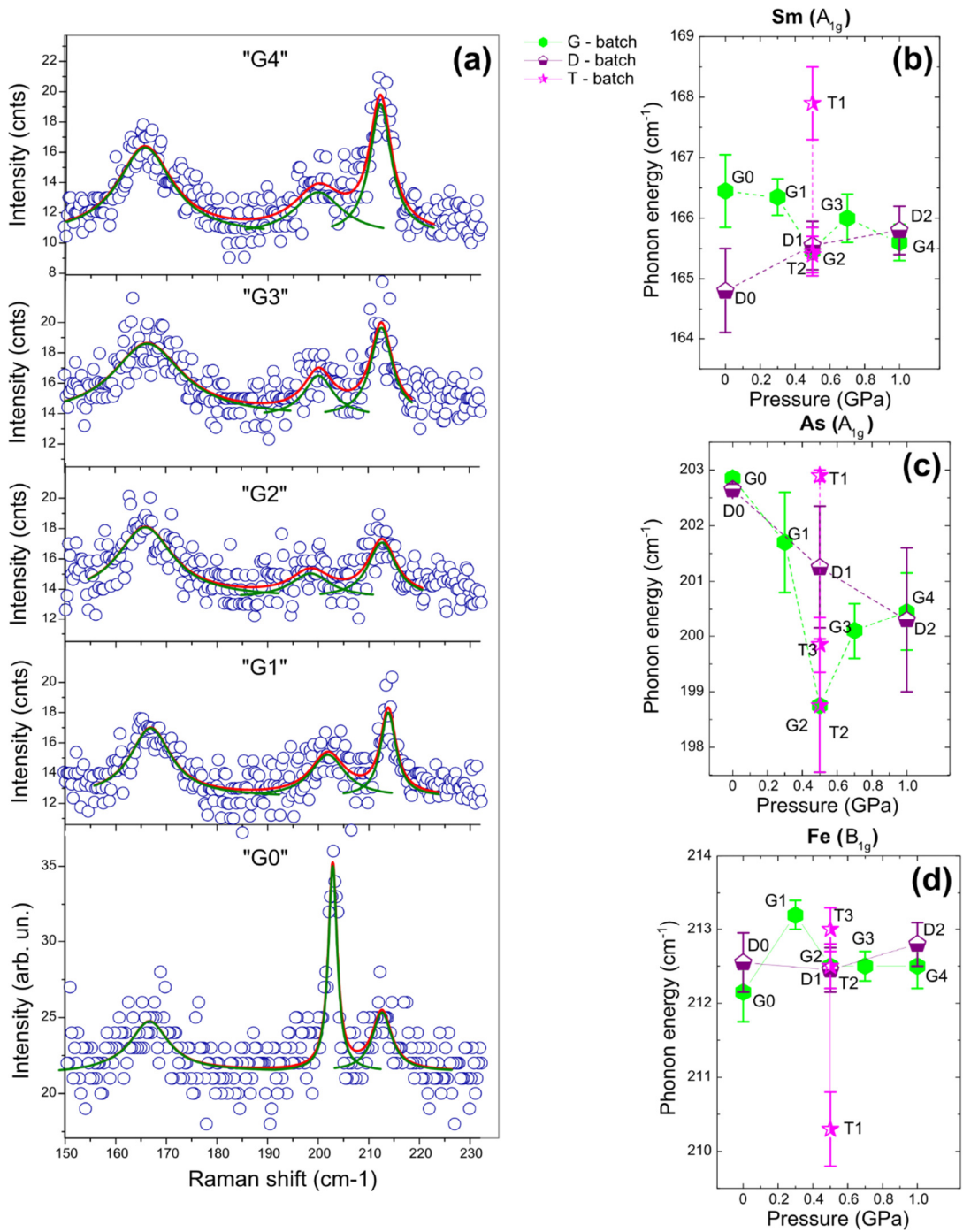

**Figure 6:** Backscattered electron image (BSE; AsB) of **(a–c)** the parent P sample, and **(d–f)** G0: 0 GPa, **(g–i)** G1: 0.3 GPa, **(j–l)** G2: 0.5 GPa and **(m–o)** G3: 0.7 GPa **(p–r)** G4: 1 GPa. Bright contrast, light grey, and black contrast correspond to the phases of $Sm_2O_3$, $SmFeAsO_{0.8}F_{0.2}$, and pores. The black contrast can be SmAs/FeAs at few places.

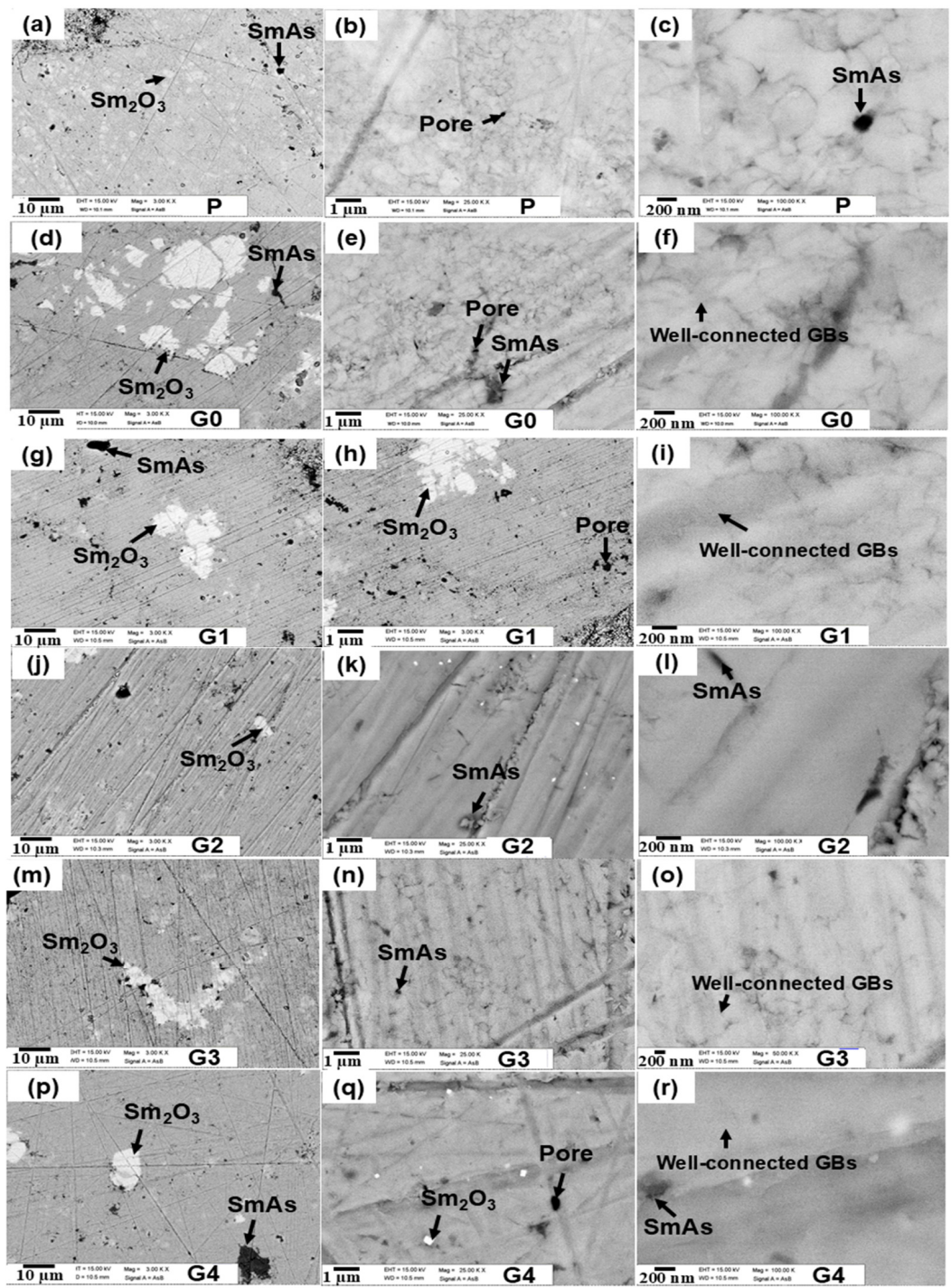

**Figure 7:** **(a)** The variation of resistivity ($\rho$) with the temperature up to the room temperature **(b)** Low-temperature variation of the resistivity up to 60 K for the parent P sample synthesised under various pressure 0-1 GPa prepared by HP-HTS method, and ground and pelletized samples: G0: 0 GPa, G1: 0.3 GPa, G2: 0.5 GPa and G3: 0.7 GPa and G4: 1 GPa.

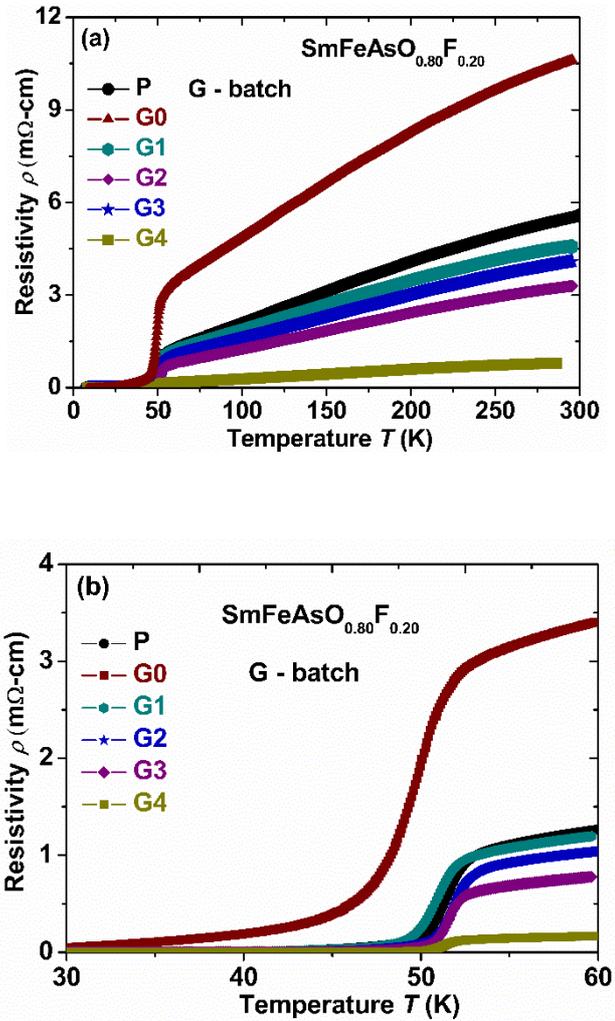

**Figure 8: (a)** The variation of resistivity ($\rho$) with the temperature up to the room temperature **(b)** Low-temperature variation of the resistivity up to 60 K for the parent P sample, and direct pellet samples D0: 0 GPa, D1: 0.5 GPa, D2: 1 GPa prepared by HP-HTS.

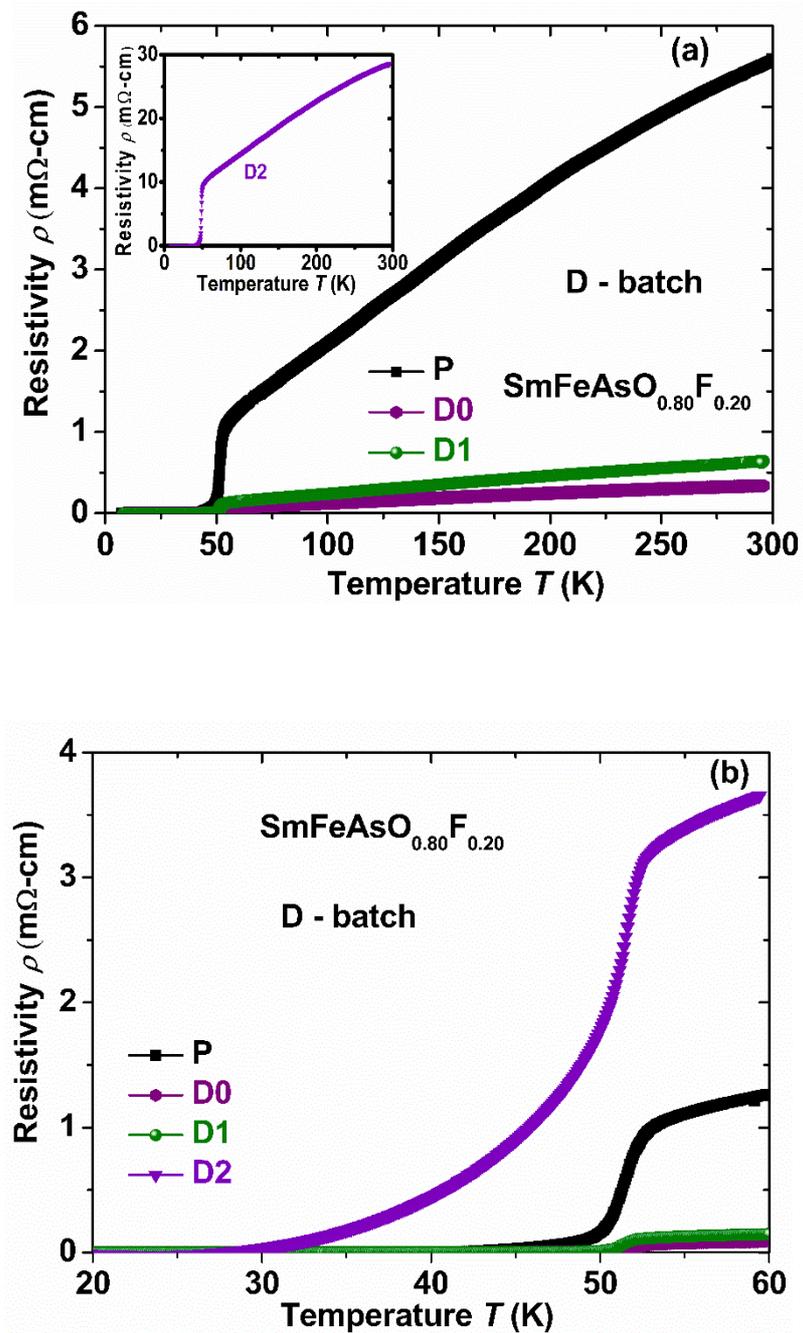

**Figure 9:** **(a)** The variation of resistivity (*ρ*) with the temperature up to the room temperature **(b)** Low-temperature variation of the resistivity up to 60 K for the parent P sample, and ground and pelletized parent samples for different heating times T1 for 0.5 h, T2 for 1 hour and T2 for 2 hours, prepared by HP-HTS at constant temperature 900 °C and growth pressure of 0.5 GP.

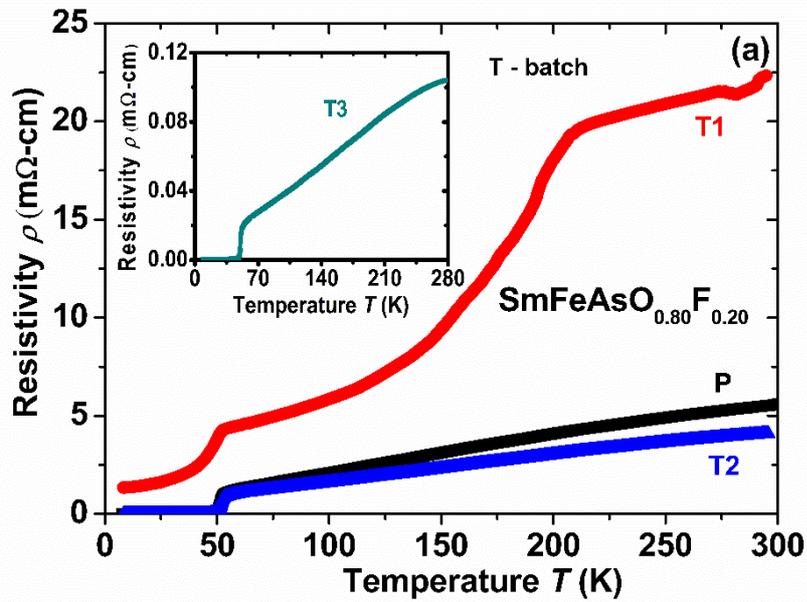

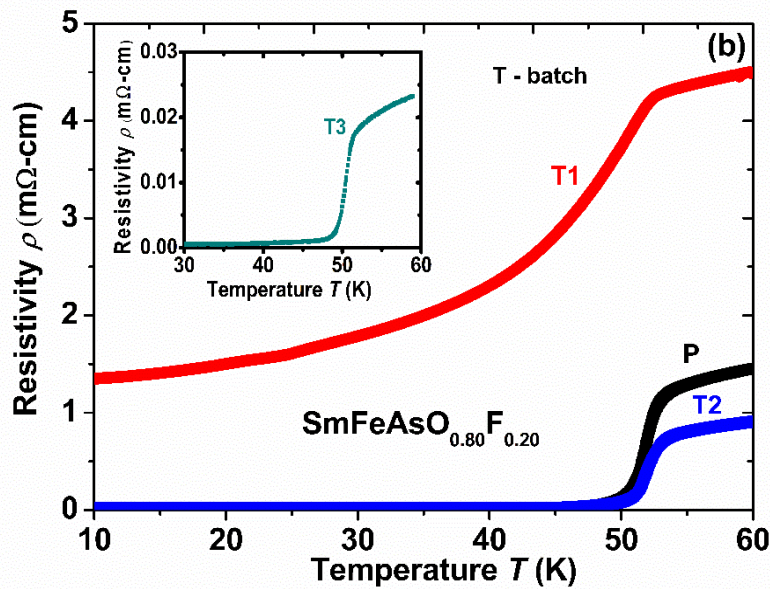

**Figure 10:** The temperature dependence of the normalized magnetic moment in ZFC and FC mode at the magnetic field of 20 Oe for **(a)** the parent P sample, and ground and pelletized samples: G2: 0.5 GPa and G4: 1 GPa with Parent P sample **(b)** the parent P sample, and direct pellet samples D0: 0 GPa, D1: 0.5 GPa, D2: 1 GPa **(c)** the parent P sample, and ground and pelletized parent samples for different heating times T1 for 0.5 h, T2 for 1 hour and T2 for 2 hours prepared by HP-HTS.

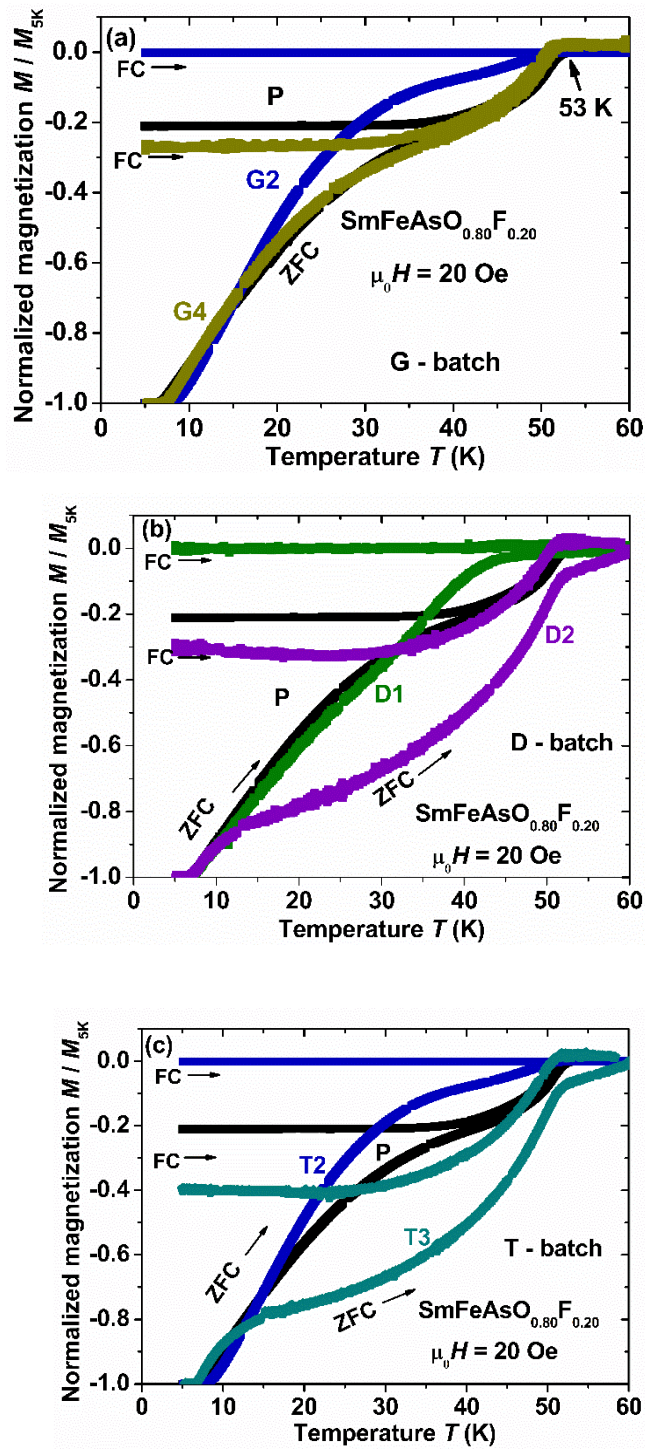

**Figure 11:** The variation of the critical current density ($J_c$) at 5 K with the applied magnetic field up to 9 T **(a)** for the parent sample, G2 and G4 at 5 K **(b)** for parent P, D1, D2 **(c)** for parent P, T2 and T3. The inset shows the magnetic hysteresis loop (*M-H*) for P, G2 and G4 in Figure (a), P, D1 D2 in Figure (b) and P, T2, T3 in Figure (c).

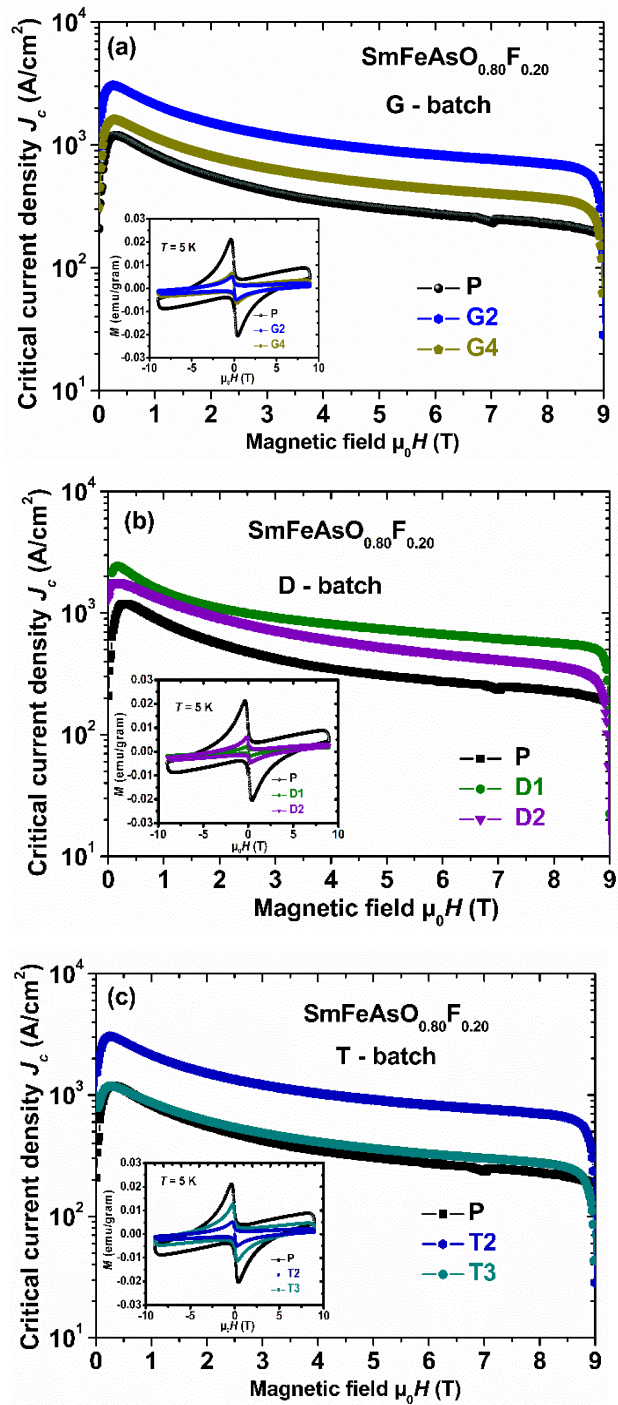

**Figure 12:** The variation of **(a)** the onset transition temperature ($T_c^{onset}$) **(b)** the transition width ($\Delta T$), **(c)** room temperature resistivity ($\rho_{300\,K}$) **(d)** RRR (= $\rho_{300\,K}$ / $\rho_{60\,K}$), **(e)** critical current density ($J_c$) of SmFeAsO$_{0.8}$F$_{0.2}$ with the various growth pressure for the parent P and G-batch: G0: 0 GPa, G1: 0.3 GPa, G2: 0.5 GPa and G3: 0.7 GPa and G4: 1 GPa prepared by HP-HTS.

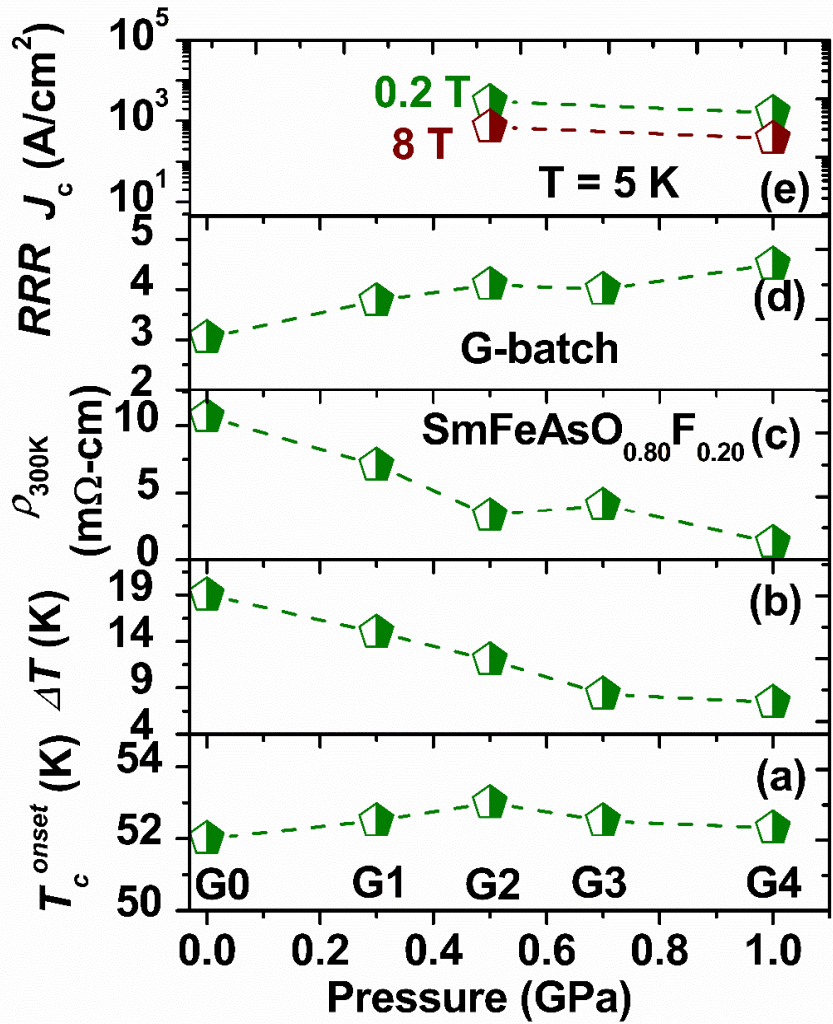

**Figure 13:** The variation of **(a)** the onset transition temperature ($T_c^{onset}$) **(b)** the transition width ($\Delta T$), **(c)** room temperature resistivity $\rho_{300\ K}$ **(d)** RRR (= $\rho_{300\ K}$ / $\rho_{60\ K}$), **(e)** critical current density ($J_c$) of SmFeAsO$_{0.8}$F$_{0.2}$ with the various growth pressure for the parent P and D-batch samples. D0, D1, and D2 are prepared by HP-HTS at constant temperature 900 °C and time 1 h under different growth pressure of 0, 0.5, and 1 GPa.

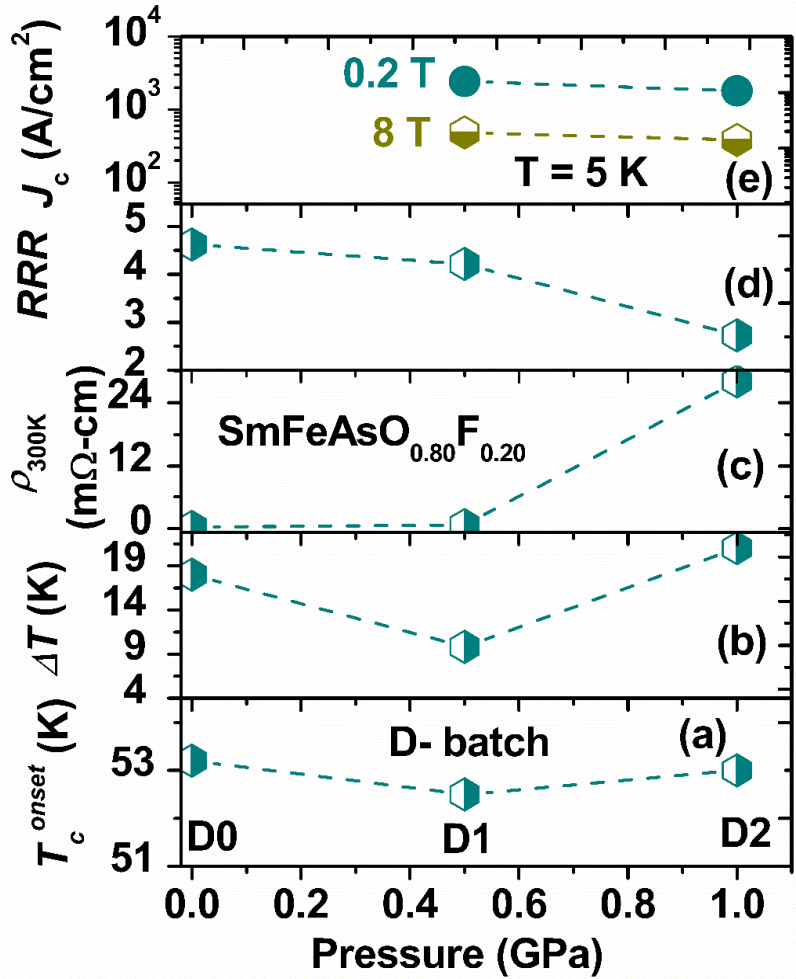

**Figure 14:** The variation of **(a)** the onset transition temperature ($T_c^{onset}$) **(b)** the transition width ($\Delta T$), **(c)** room temperature resistivity $\rho_{300\,K}$ **(d)** $RRR$ (= $\rho_{300\,K}$ / $\rho_{60\,K}$), **(e)** critical current density ($J_c$) of SmFeAsO$_{0.8}$F$_{0.2}$ with the various growth pressure for the parent P and T-batch samples. T1, T2, and T3 are prepared by HP-HTS at constant temperature 900 °C and the growth pressure of 0.5 GPa under different heating temperature of 0.5, 1 and 2 hours, respectively.

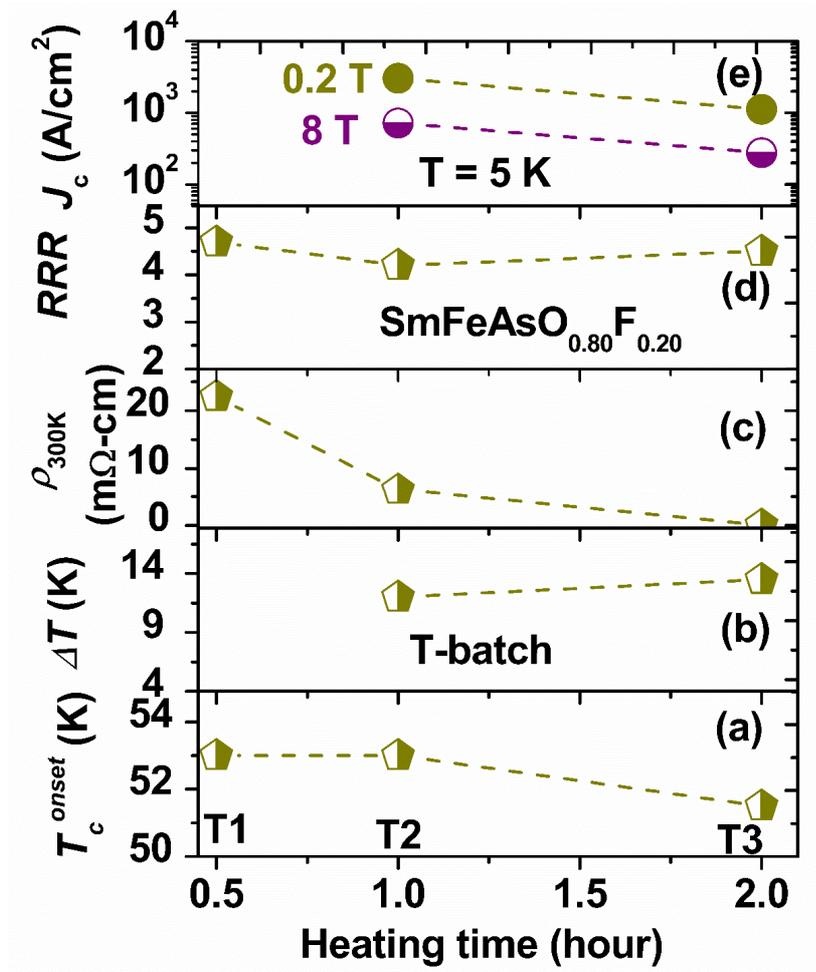

**Figure 15:** The variation of **(a)** the onset transition temperature ($T_c^{onset}$) **(b)** the critical current density $J_c$ values of various Sm1111 bulk samples prepared by HP-HTS with the growth pressure. The $J_c$ dependence of **(c)** the superconducting Sm1111 phase (%) obtained from XRD and **(d)** the sample density for some selected G, D and T-batch samples.

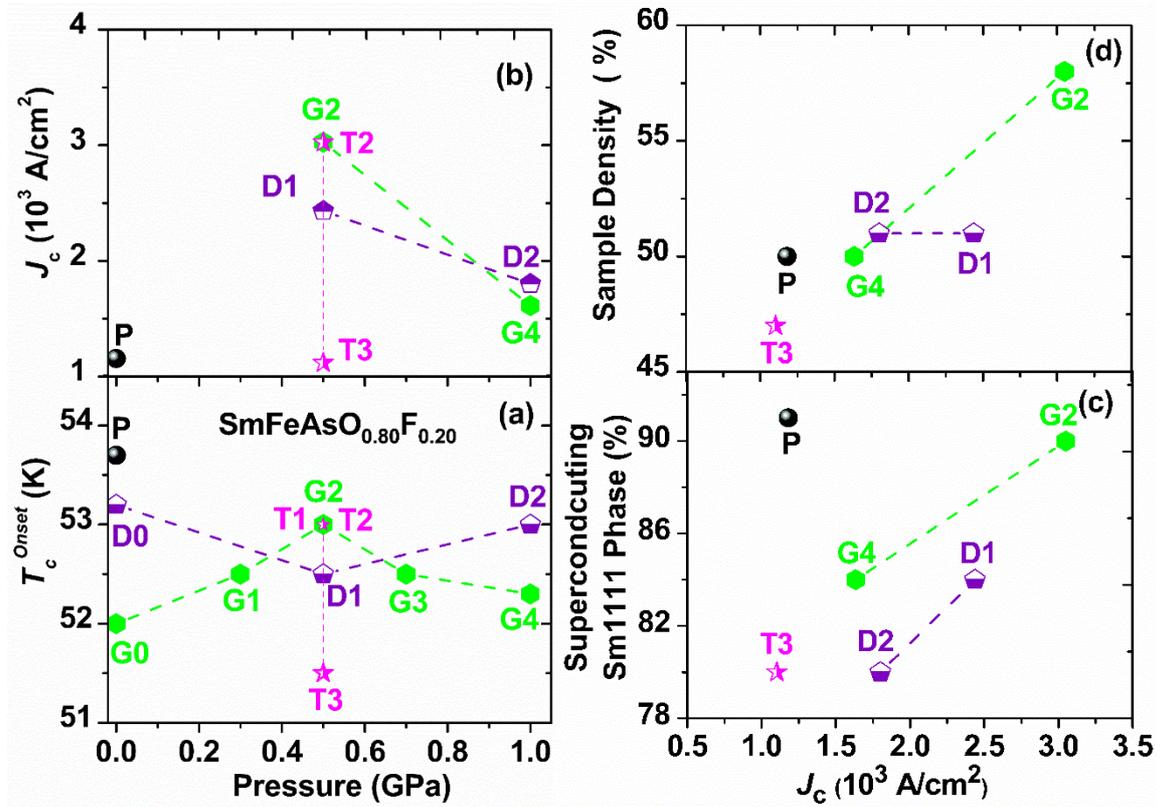

# High-pressure growth effects on the superconducting properties of Sm-based oxypnictide superconductors

Mohammad Azam[1], Manasa Manasa[1], Tatiana Zajarniuk[2], Ryszard Diduszko[3, 4], Taras Palasyuk[1], Tomasz Cetner[1], Andrzej Morawski[1], Cezariusz Jastrzębski[5], Andrzej Szewczyk[2], Michał Wierzbicki[5], Shiv J. Singh[1,††]

[1]*Institute of High Pressure Physics (IHPP), Polish Academy of Sciences, Sokołowska 29/37, 01-142 Warsaw, Poland*

[2]*Institute of Physics, Polish Academy of Sciences, Aleja Lotników 32/46, 02-668 Warsaw, Poland*

[3]*Łukasiewicz Research Network Institute of Microelectronics and Photonics, Aleja Lotników 32/46, 02-668 Warsaw, Poland*

[4]*National Center for Nuclear Research, ul. Andrzeja Sołtana 7, 05-400 Otwock, Świerk, Poland*

[5]*Faculty of Physics, Warsaw University of Technology, Koszykowa 75, 00-662 Warsaw, Poland*

[††]Corresponding author: Shiv J. Singh

Email: sjs@unipress.waw.pl

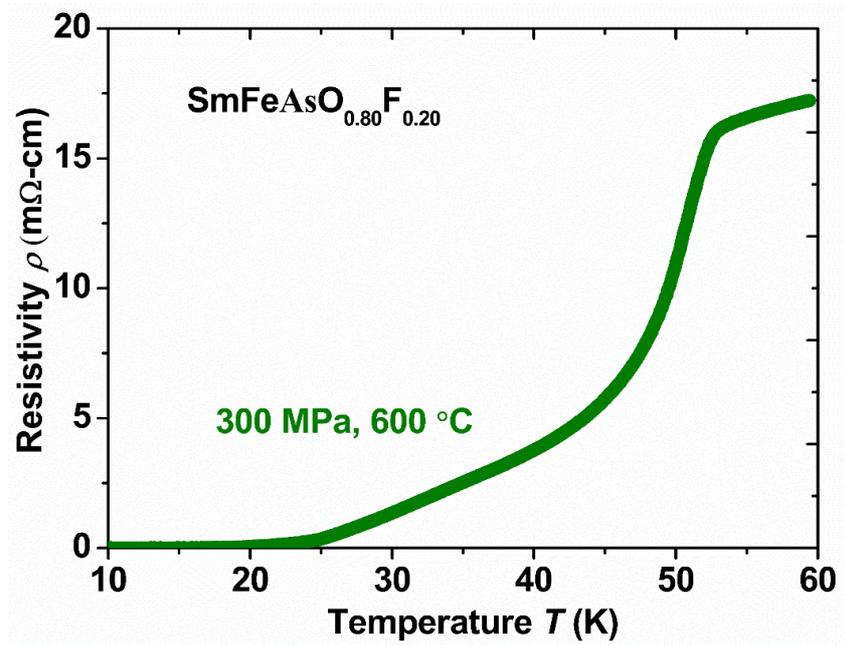

**Figure S1.** The temperature dependence of resistivity behaviour of SmFeAsO$_{0.80}$F$_{0.20}$ bulks synthesized by HP-HTS technique prepared at 0.3 GPa for 1 h at 600 °C.

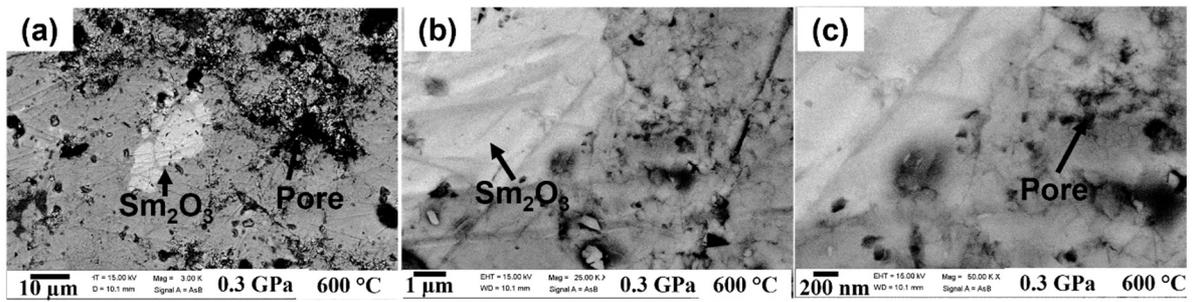

**Figure S2.** Backscattered electron image (BSE; AsB) of SmFeAsO$_{0.80}$F$_{0.20}$ bulks synthesized by HP-HTS technique prepared at 0.3 GPa for 1 h at 600 °C.

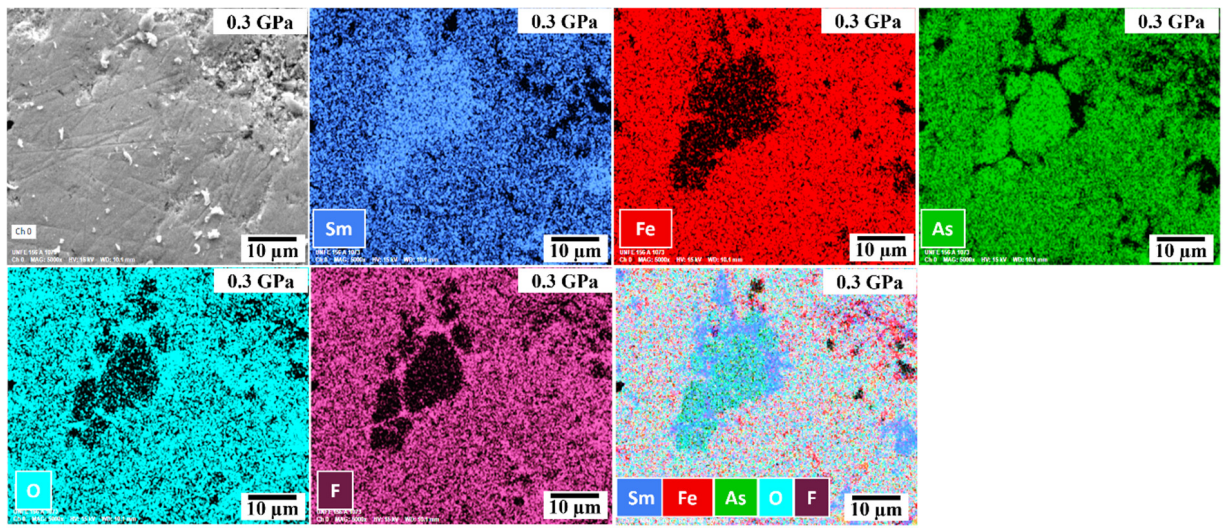

**Figure S3.** Mapping of all constituent elements of SmFeAsO$_{0.80}$F$_{0.20}$ bulks synthesized by HP-HTS technique prepared at 0.3 GPa for 1 h at 600 °C.

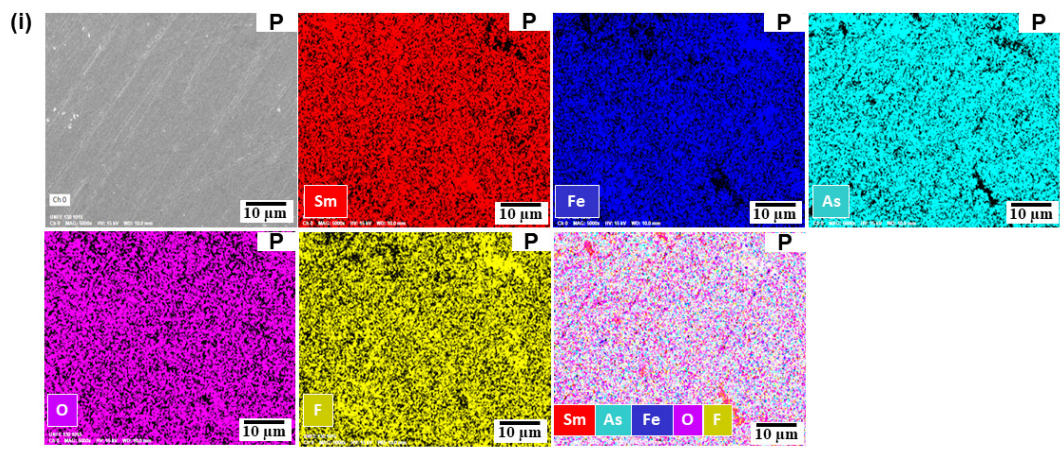
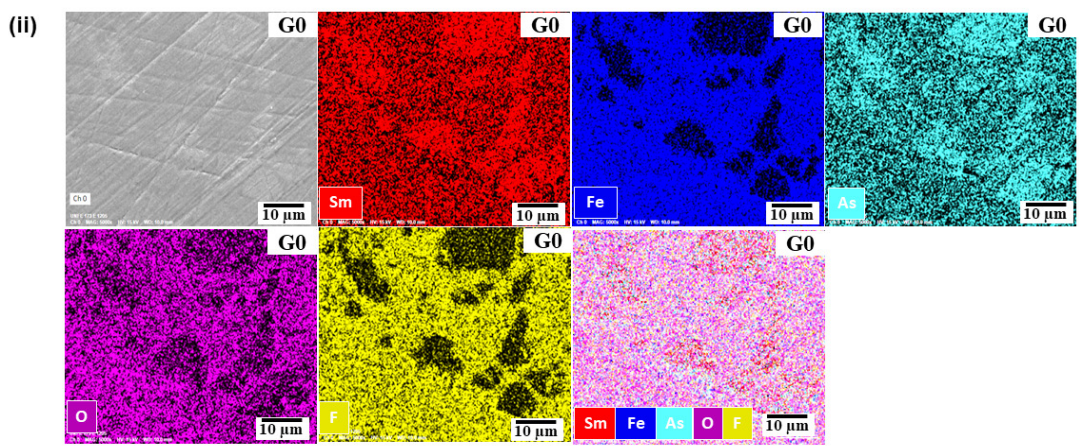
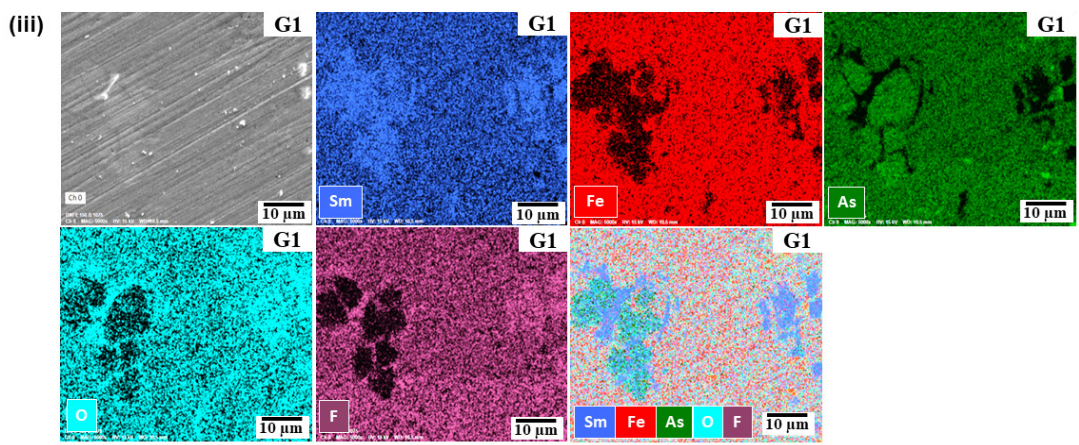

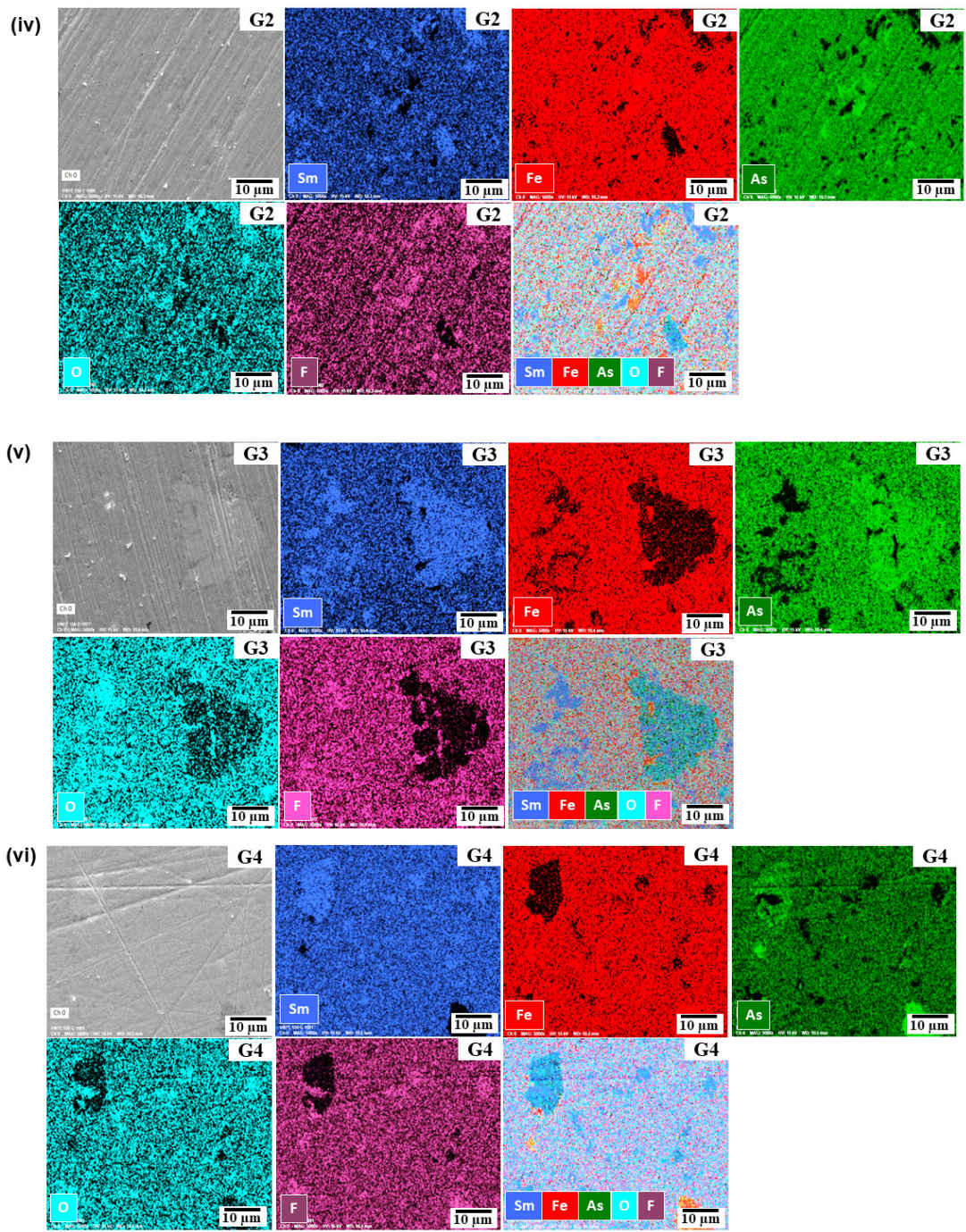

**Figure S4:** Elemental mapping for the constituent elements of SmFeAsO$_{0.8}$F$_{0.2}$ polycrystalline samples: **(i)** Parent P, and and ground and pelletized samples **(ii)** G0, **(iii)** G1, **(iv)** G2, **(v)** G3 and **(vi)** G4 prepared by HP-HTS at constant temperature 900 °C and time 1 h under different growth pressures of 0, 0.3, 0.5, 0.7, and 1 GPa, respectively. The first and last images in each sample series are SEM images, while the final image is a combined image of all the constituent elements. The rest of the images depict the elemental mapping of an individual element: Sm, Fe, As, O, and F.

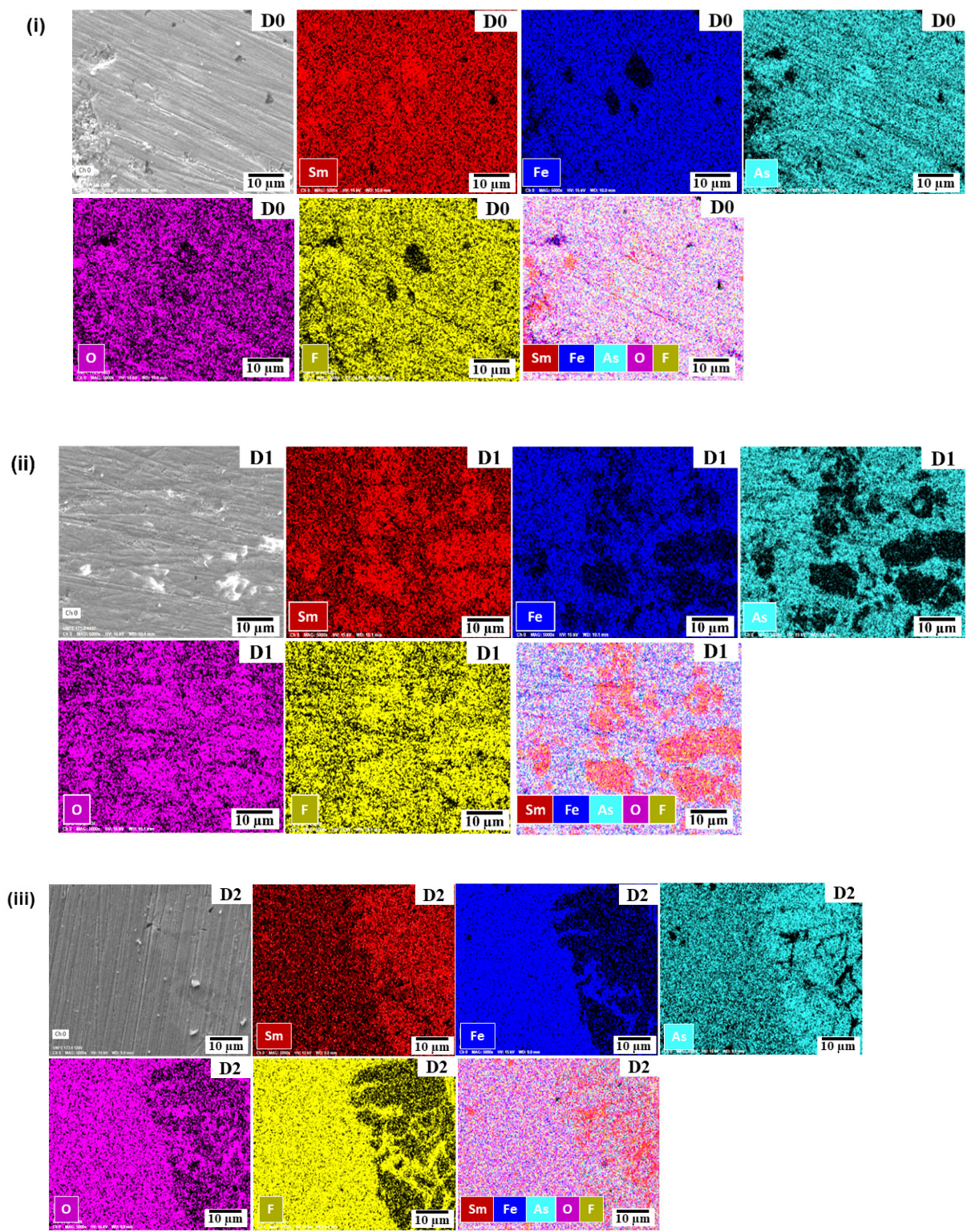

**Figure S5:** Elemental mapping for the constituent elements of SmFeAsO$_{0.8}$F$_{0.2}$ polycrystalline samples: **(i)** D0, **(ii)** D1 **(iii)** D2, prepared by HP-HTS at constant temperature 900 °C and time 1 h under different growth pressures of 0 GPa, 0.5 GPa and 1 GPa, respectively. The first and last images of each sample series are SEM images and a combined image of all the constituent elements, respectively. The rest images depict the elemental mapping of an individual element Sm, Fe, As, O and F.

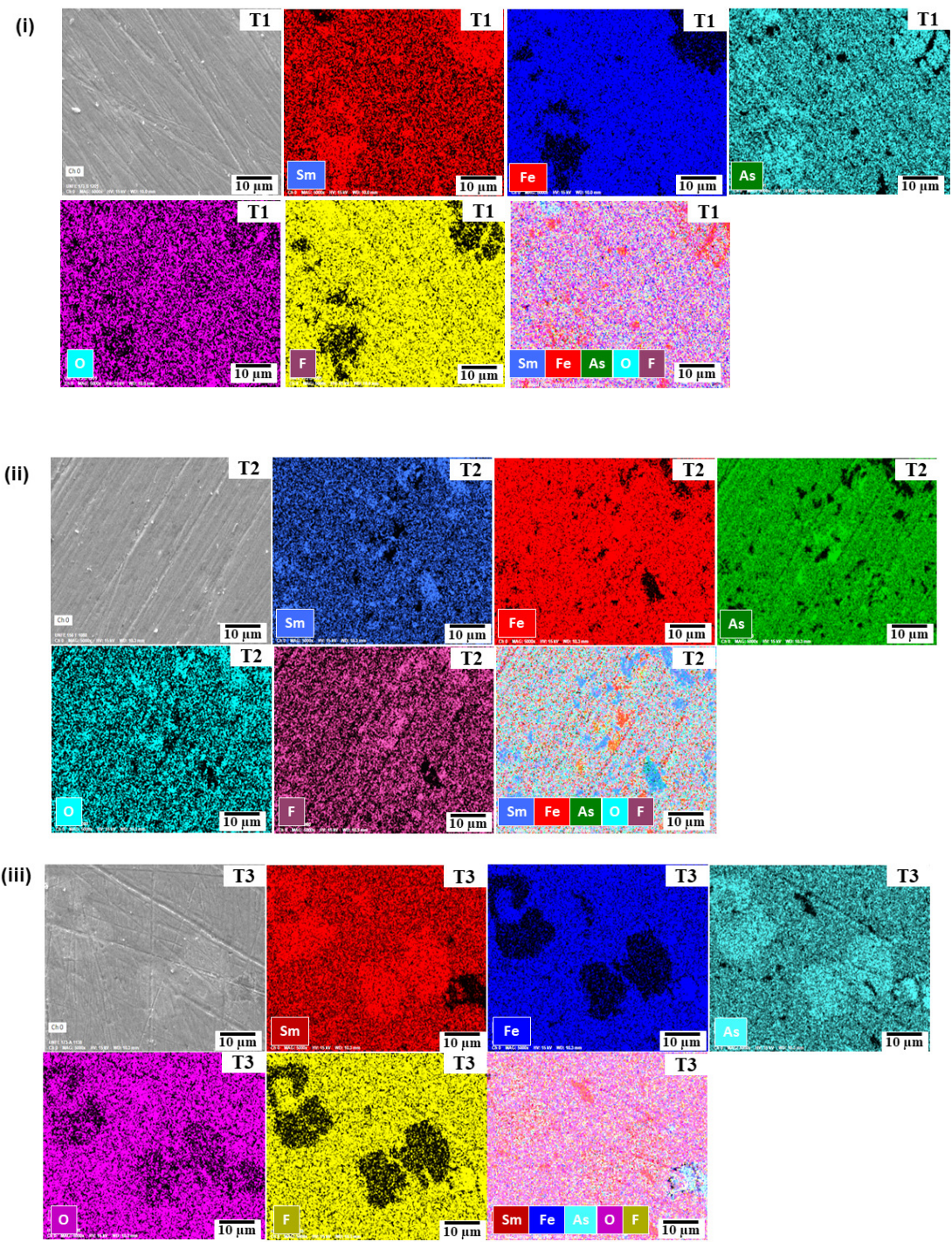

**Figure S6:** Elemental mapping for the constituent elements of T-batch bulks **(i)** T1 for 0.5 h, **(ii)** T2 for 1 hour and **(iii)** T3 for 2 hours, prepared by HP-HTS at constant temperature 900 °C and time 1 h under a growth pressure of 0.5 GPa. The first and last images of each sample series are SEM image and a combined image of all the constituent elements, respectively. The rest of the images depict the elemental mapping of individual elements Sm, Fe, As, O and F. The sample T2 has the same growth condition as the G2 sample

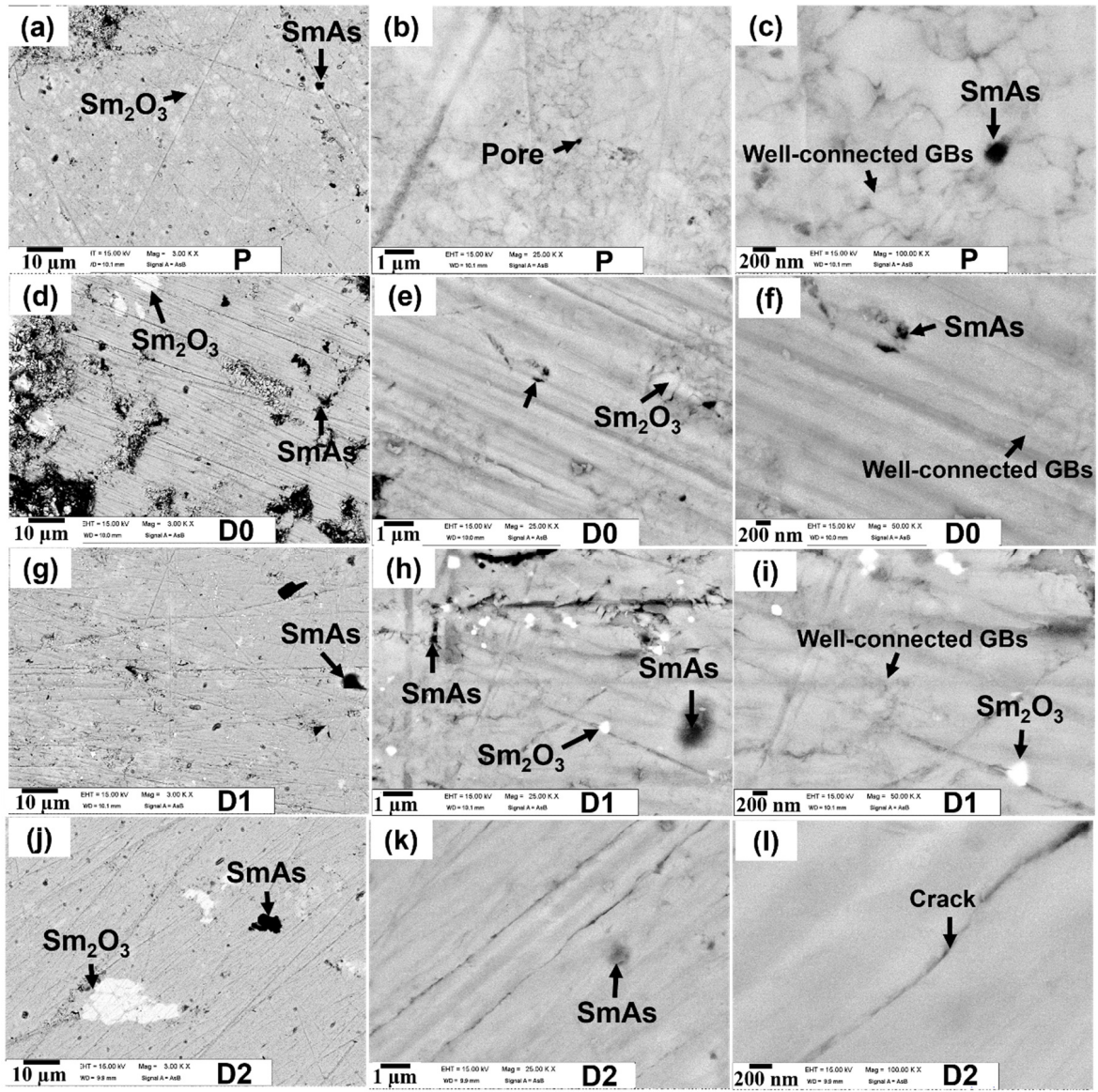

**Figure S7:** Backscattered electron image (BSE; AsB) of **(a–c)** the parent P sample, and **(d–f)** D0: 0 GPa, **(g–i)** D1: 0.5 GPa, **(j–l)** D2: 1 GPa. Bright contrast, light grey, and black contrast correspond to the phases of $Sm_2O_3$, $SmFeAsO_{0.8}F_{0.2}$, and pores. The black contrast can be SmAs/FeAs at few places.

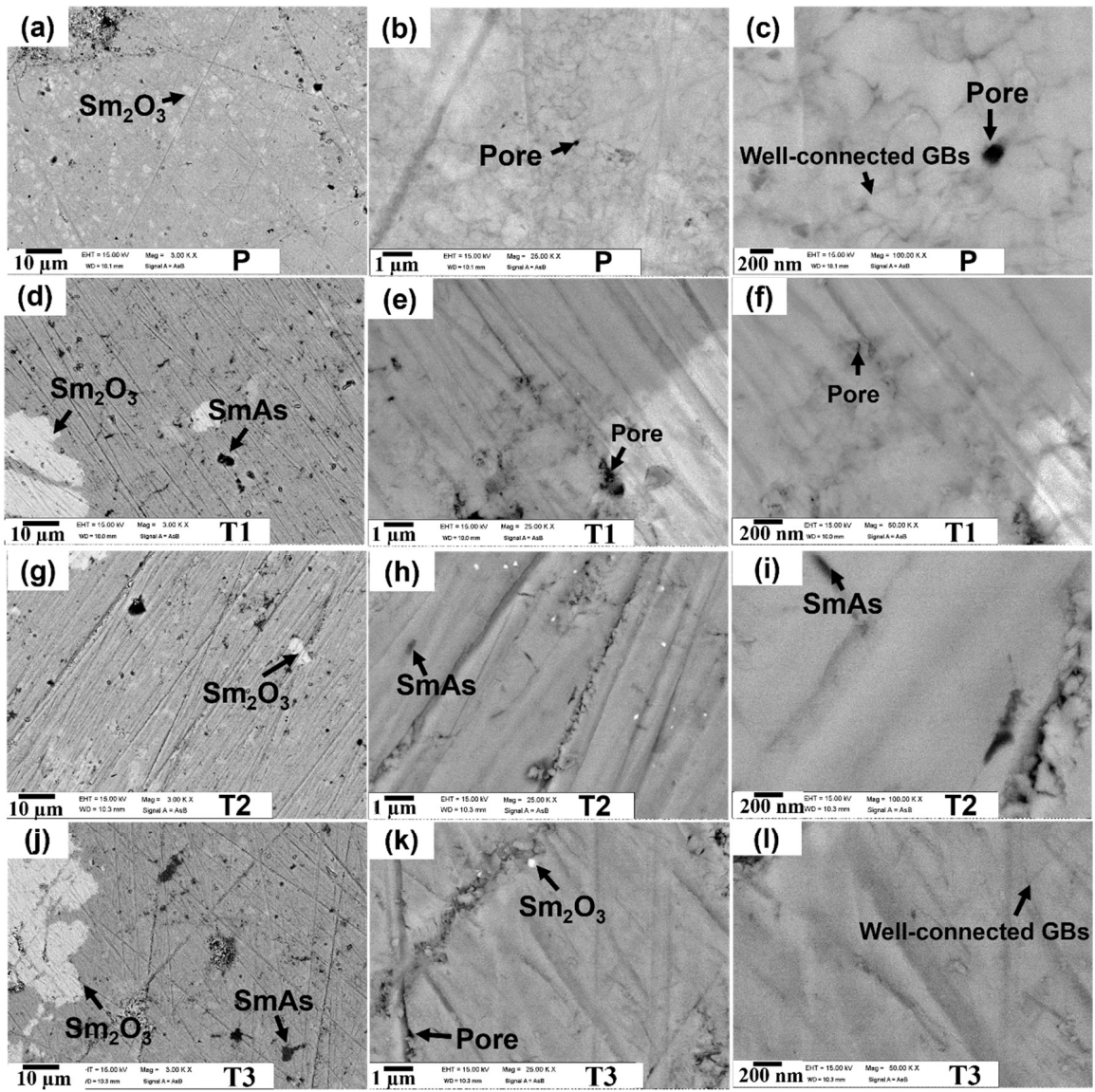

**Figure S8:** Backscattered electron image (BSE; AsB) of **(a–c)** the parent P sample, and **(d–f)** T1 for 0.5 h, **(g–i)** T2 for 1 hour and **(j–l)** T3 for 2 hours, prepared by HP-HTS at constant temperature 900 °C and time 1 h under a growth pressure of 0.5 GP. Bright contrast, light grey, and black contrast correspond to the phases of $Sm_2O_3$, $SmFeAsO_{0.8}F_{0.2}$, and pores. The black contrast can be SmAs/FeAs at few places.

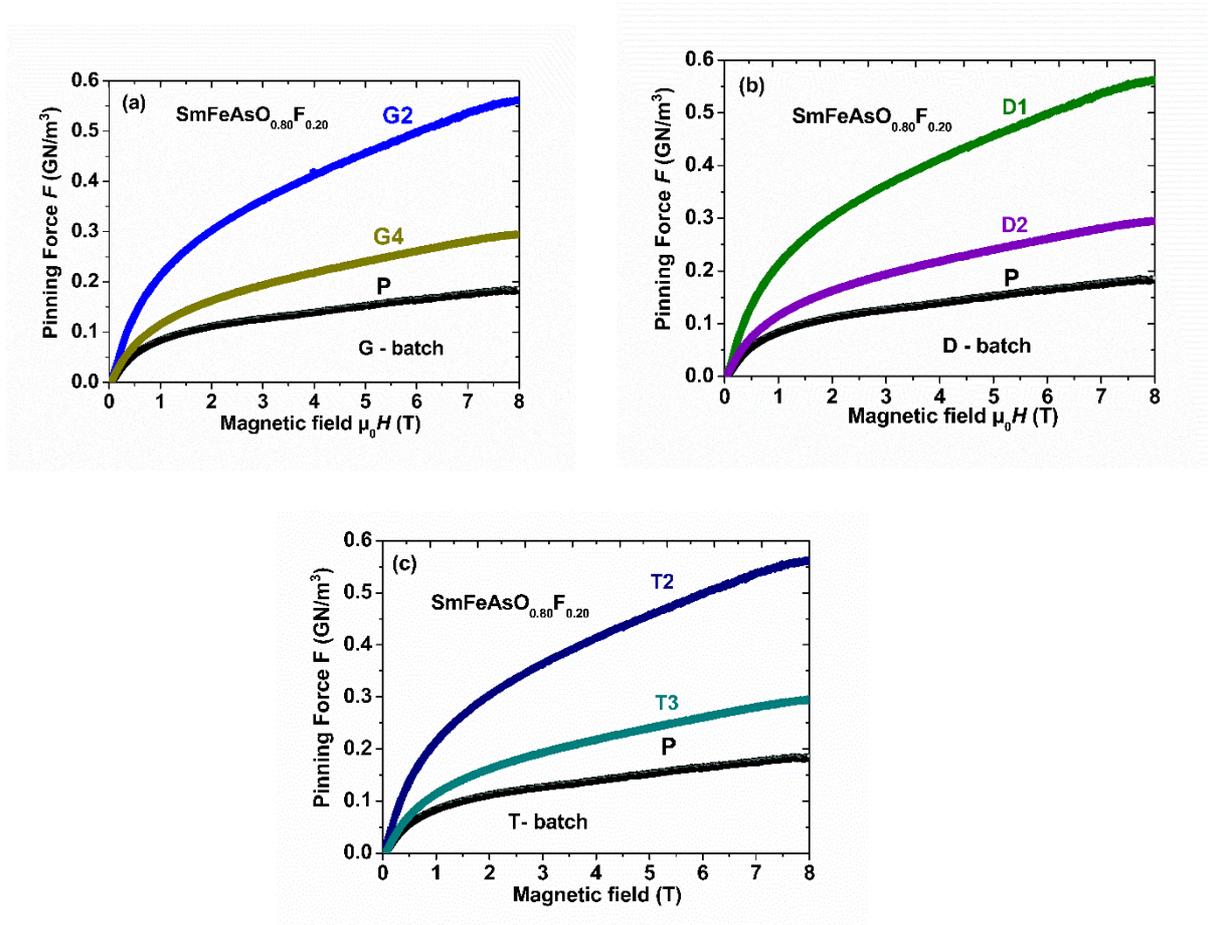

**Figure S9:** The pinning force at 5 K with the applied magnetic field up to 8 T for (a) the parent P sample, and ground and pelletized samples: G2: 0.5 GPa and G4: 1 GPa. (b) the parent P sample, and direct pellet samples D1: 0.5 GPa, D2: 1 GPa. (b) the parent P sample, and ground and pelletized parent samples for different heating times T2 for 1 h, and T3 for 2 hours prepared by HP-HTS.